\documentclass[journall=acsjctc,manuscript=article]{achemso}
\usepackage{amsmath}
\usepackage{color}
\usepackage{amssymb}
\usepackage{graphicx}
\usepackage{blindtext}
\usepackage{multirow}
\usepackage{booktabs}
\usepackage{chngcntr}
\usepackage{subfigure}
\usepackage[table,xcdraw]{xcolor}
\usepackage{booktabs}
\usepackage{hhline}
\usepackage{gensymb}

\usepackage{float}

\def\MAl{Mn$_{4}$Al$_{11}$}
\def\MGe{Mn$_{4}$Al$_{9}$Ge$_{2}$}
\def\MG{Mn$_{4}$Al$_{10}$Ge}
\def\etal{{\it et al. }}

\title{Doping Induced Magnetic and Electronic phase Transition in Ferrimagnetic Half-metallic  Mn$_{4}$Al$_{11}$ Compound}
\author{Sujoy Datta}
\email{sujoydatta13@gmail.com}
\affiliation{Kadihati KNM High School, Ganti, Kolkata 700132, India.}

\author{Prashant Singh}
\email{psingh84@ameslab.gov}
\affiliation{Ames National Laboratory, U.S. Department of Energy, Iowa State University,  Ames, Iowa 50011, USA.}


\begin{tocentry}
	\includegraphics[width=\columnwidth]{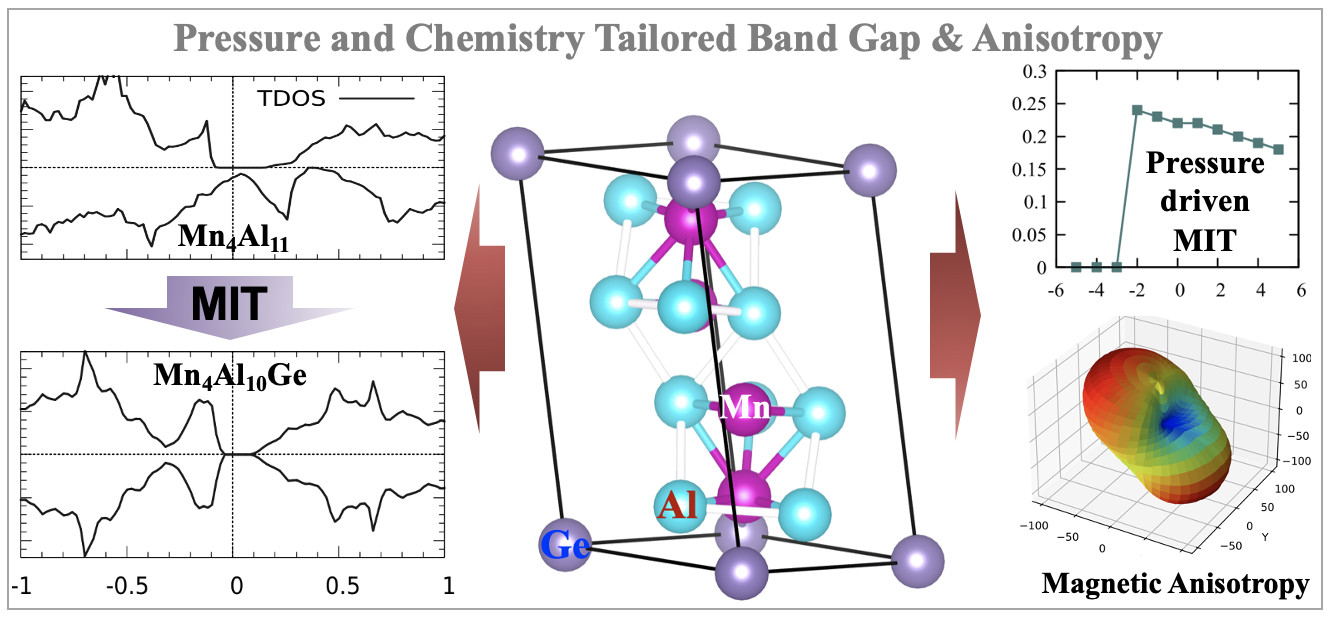}
\end{tocentry}

\begin{document}
	
\begin{abstract}
The future of spintronic and semiconductor applications demands materials with tailored electronic and magnetic properties. This study uses density functional theory to investigate the electronic structure of the half-metallic compound Mn$_{4}$Al$_{11}$ under uniaxial strain and in its Ge-substituted derivatives. Strain analysis shows that although the half-metallic band-gap collapses under strain beyond $-2\%$, the ferrimagnetic character remains stable. Ge substitution at six inequivalent Al-sites in Mn$_{4}$Al$_{11}$ results in varying degrees of metallicity and magnetic properties. Substitution at Al$=(000)$ induces a metal-to-insulator transition with an indirect semiconducting gap of $0.14~ eV$. Bonding and hybridization analysis reveals that local Mn-Al interactions due to Ge substitution significantly modify the local electronic structure, causing both electronic and magnetic phase transitions. This work highlights the effectiveness of substitutional doping in tuning half-metallicity and magnetic properties in inorganic solids, enabling the design of materials for future technological applications.
\end{abstract}

	
\section{Introduction}
The dawn of a new era in spintronic, semiconductor, catalysis, and battery technologies hinges on the availability of a plethora of materials with precisely tailored electronic and magnetic properties. Intermetallic compounds, formed from two or more metals, are ubiquitous in this field, offering a prolific array of magnetic and electronic behaviors \cite{stoloff2000emerging, zhang2024recent}. These materials possess remarkable versatility, exhibiting properties like half-metallicity, ferromagnetism, ferrimagnetism, and tunable conductivity \cite{terada2002thermal,berry2017enhancing}. Their ability to precisely control both electronic structure and magnetic moments positions them as key candidates for next-generation technologies such as quantum computing, energy-efficient devices, and advanced data storage \cite{yuan2023honeycomb,vaney2022topotactic,adamski2024selective}. With atomic-level manipulation, intermetallic compounds are poised to become indispensable in shaping the future of high-performance spintronic and semiconductor materials.

{\par}{Among various intermetallic systems, Manganese-Aluminium (Mn-Al) based compounds have attracted particular interest due to their unique combination of structural and functional properties. The electronic structure of Mn-Al based intermetallic compounds plays a crucial role in their mechanical and electrical properties  \cite{polmear2017light}. The interaction between Mn and Al atoms, particularly in alloys like Mn$_6$Al and Mn$_{12}$Al, influences their strength, ductility, and conductivity \cite{wei2009}. In such alloys, the Manganese atoms play a crucial role in controlling the structural rigidity. Manganese's 3$d$ orbitals create localized electronic states that modify the density of states (DOS) near the Fermi level (E$_F$), contributing to the formation of Mn$_6$Al microstructures that enhance tensile strength  \cite{nam2000effect}. Besides, the bonding driven by the hybridization of Mn-3$d$ and Al-3$p$ orbitals, strengthens the material's stability and hardness.} Additionally, the Mn-Al alloys exhibit tunable magnetic properties, ranging from ferromagnetic to ferrimagnetic behavior, depending on Mn content and phase. This tunability allows for precise control of mechanical, electrical, and magnetic properties, making Mn-Al intermetallics suitable for advanced applications in aerospace, automotive, and spintronics \cite{shen2023mn}.

\begin{figure*}[t]
	\centering	
	\includegraphics[width=\columnwidth]{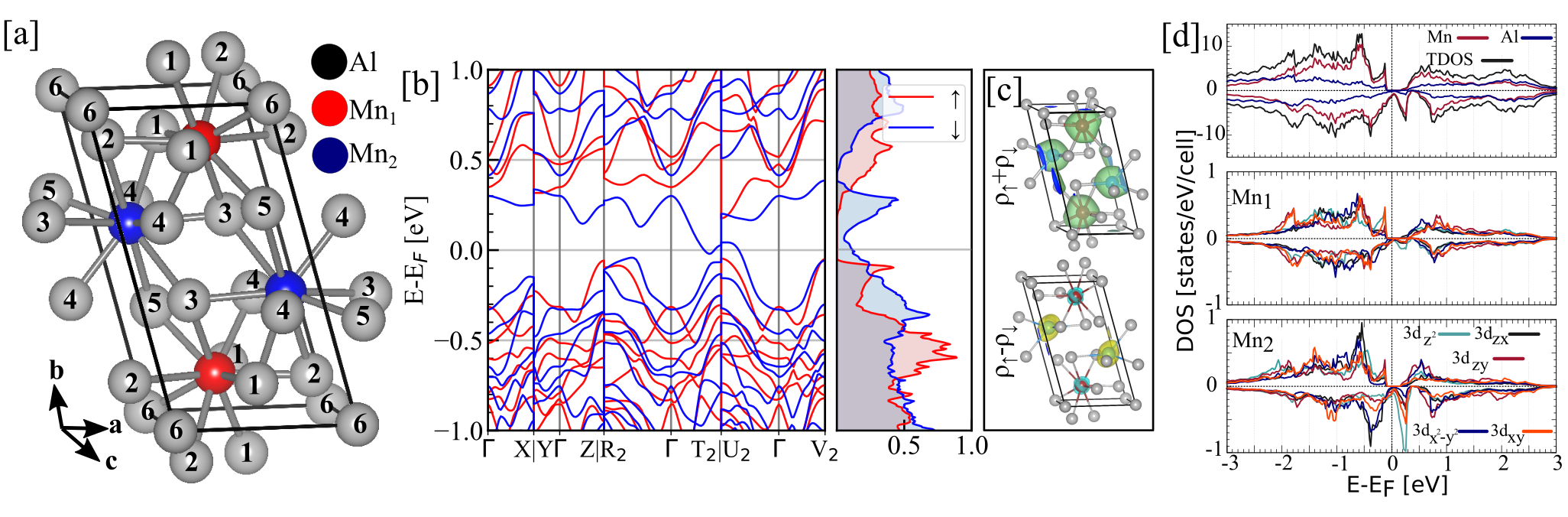}
	\caption{\label{fig:MnAl} (a) Crystal structure of \MAl; (b) Spin resolved energy band dispersion and density of state; (c) Charge density (total and difference of up and down) of \MAl; (d) Spin resolved partial density of state of \MAl.}
\end{figure*}

{\par} {Over the years, extensive research has been conducted on Mn-Al alloy systems \cite{mcalister1987mn,liu1999thermodynamic,shen2023mn,gorshenkov2020magnetic,du2007reassessment}. Notably, McAlister and Murray's comprehensive phase diagram reveals a rich assortment of intermetallic phases within this system \cite{mcalister1987mn}, with phase formation extending up to approximately $22\%$ Mn concentration \cite{murray1987stable}.  These phases include orthorhombic MnAl$_6$, triclinic MnAl, the metastable hexagonal M$_3$Al$_{10}$, and two hexagonal approximants, $\mu$-MnAl$_4$ and $\lambda$-MnAl$_4$ \cite{murray1987stable,shukla2009thermodynamic}. The $\{\mu, \lambda\}$ phases of MnAl$_4$ exhibit structural features similar to those of quasicrystals.}
Notably, the \MAl~ structure displays a layered arrangement resembling the Penrose lattice \cite{kumar1986}, suggesting its classification as an imperfect, decagonal quasicrystalline alloy \cite{mihalkovivc1996}. At high temperatures ({$\gtrsim  923 ^{\circ}~C$}), an additional \MAl~ orthorhombic polymorph has been confirmed, expanding the complexity of Mn-Al phase behavior and its structural diversity at varying conditions, as seen in numerous experimental studies \cite{kontio1980,bland1958,schaefer1986formation,murray1987stable,godecke1971supplement,grushko2008study}.
	
{\par} {Manganese, a material with variable valency and a magnetic character strongly influenced by its local environment, has long been a cornerstone in the development of alloys with remarkable electronic and magnetic properties—particularly within the Mn-Al system \cite{caplin1968,bak1985,zhao2015,aghili2016half,yao2016first}. By varying the concentration of Mn, the electronic and magnetic properties of Mn-Al alloys can be significantly tuned. For instance, the ferromagnetic phase observed in Mn-rich Mn-Al alloys ($50-75\%$ Mn) is primarily attributed to the formation of the $\tau$-phase, which crystallizes in an orthorhombic structure (space group: P4/mmm) \cite{bohlmann1981,kono1958,van1995}.} This phase, analogous to CuAu, is characterized by strong magneto-crystalline anisotropy and exhibits an orbital two-channel Kondo effect, contributing to its unique magnetic behavior \cite{zhao2015,zhu2016,sakuma1994}. However, the magnetic properties of Mn in other Mn-Al alloys are not straightforward and are heavily influenced by the local coordination environment around the Mn atoms. The magnetic moment of Mn is not only determined by its Pauling valence but also by the oscillatory nature of the magnetic exchange interactions, which vary with Mn-Mn distances \cite{mori1968localized}. These interactions can shift between positive and negative values, further complicating the understanding of the system's overall magnetic behavior \cite{price1978indirect}. Additionally, Mn alloys with Aluminium or Aluminium-Carbon combinations are known to produce permanent magnets, but the magnetic properties of Mn-Al alloys are highly process-dependent, with variations seen in different synthesis conditions \cite{park2010, kuo1992}.

{\par} {While varying the concentration of constituents is a well-established method for modifying the electronic and magnetic properties of alloys, substitutional doping is often preferred in intermetallic systems to achieve such tunability. Mn and Al have a small difference in electronegativity with values of $\chi = 1.75$ and $\chi = 1.61$ on the Allen scale, respectively. To enhance charge transfer and thereby strengthen bonding, substituting Al with elements of higher electronegativity, such as Si ($\chi = 1.916$) or Ge ($\chi = 1.994$), presents a promising strategy for better control over the electronic and magnetic behavior of Mn-Al systems. The doping or substitution by Group IV elements  in Mn-Al compounds specially Ge has been recognized as an effective way of tuning phase stability, magnetic, and structural properties. For example, several Mn-Al compounds have been synthesized with Ge doping that show modified the lattice structure (structural stability) and magnetic anisotropy compared to binary Mn-Al systems \cite{tsai1988,ido1984,lalla1992,schaefer1986,sarvar2024}.}
	
{\par} This study uses first-principles density functional theory (DFT) to understand the effects of uniaxial strain and site-specific Germanium (Ge) substitution on the electronic and magnetic properties of the ferrimagnetic half-metallic compound \MAl. While uniaxial strain collapses the half-metallic band-gap beyond $-2\%$, the ferrimagnetic order remains robust, highlighting the stability of Mn-driven spin polarization under mechanical perturbations. Ge substitution at symmetry-inequivalent Al sites results in pronounced changes to the electronic and magnetic properties, particularly at the Al$= (000)$ site, induces a metal-to-insulator transition (MIT) with an indirect band-gap of $0.14~ eV$, driven by localized charge redistribution and orbital hybridization. The crystal orbital Hamiltonian populations (COHP) and Bader charge analyses reveal critical changes in bonding interactions and valence states, showcasing the strong tunability of \MAl’s properties. These findings offer valuable insights for designing materials with tailored functionalities for spintronics and low-power electronics. 

\section{Methods}
{\par}For solids, the slowly varying estimation of charge density within generalised gradient approximation (GGA) scheme  is more suitable than the atomic or molecular systems. That is why Perdew-Burke-Ernzerhof (PBE) correction modified for solids (PBEsol) is always a good choice for structural prediction of solids \cite{PBEsol}. {Projected augmented wave (PAW) basis and PBEsol exchange-correlation (xc) functional are used in all calculations done using Quantum Espresso suite \cite{QE}.} $6\times 4 \times6$ k-space mesh, Marzari-Vanderbilt cold smearing \cite{marzari1999} of width $0.02~Ry$ is used throughout along with $55~Ry$ and $400~Ry$ kinetic-energy cut-off for wavefunction and charge density, respectively. For structural optimization the convergence threshold of forces for ionic minimization is chosen as $10^{-6}~Ry$. Broyden–Fletcher–Goldfarb–Shanno (BFGS) of ionic dynamics is used.
{Structural optimizations are performed for \MAl, \MAl~ under isotropic pressure, \MGe-1 to \MGe-5, and \MG~ structures. Ionic relaxations are carried out for the structures under unidirectional stress. Total energies are calculated through separate self-consistent field (SCF) calculations. The band structure and density of states are computed using the optimized structures. The convergence threshold for self-consistency is set to $10^{-6}~Ry$. For magneto-crystalline anisotropy, spin-orbit coupling (SOC) is applied to the optimized structure.}

{\par}{Phonon dispersions are computed using the quasi-harmonic approximation via the Phonopy code, employing a finite-displacement method with a fixed displacement of $0.02~ \text{\AA}$ \cite{togo2023first}. To obtain the harmonic phonon frequencies and their normalized eigenvectors, the force-constant matrix is derived using density functional perturbation theory (DFPT) \cite{baroni2001phonons}.}

{\par} For pre- and post-processing, Vesta and ElATools are employed \cite{elatools,vesta}. The analysis of chemical bonding is conducted using the Lobster package, specifically for the PAW basis \cite{nelson2020}.

{\par}The formation energy of the Mn-Al-Ge systems is found from the formula:
\begin{eqnarray}
	E_{formation} = \frac{1}{n} [ E_{total}^{Mn-Al-Ge} - \sum_{i} n_i E_i ] 
\end{eqnarray}
where, $E_{total}^{Mn-Al-Ge}$ is the total energy of Mn-Al-Ge system and $E_i$ are the elemental energy of Mn, Al or Ge in their ground state and $n_i$ are the number of each type of atom.

\section{Results}
\subsection{Properties of Mn$_{4}$Al$_{11}$}
\subsubsection{Structure}
The structural properties of \MAl~ have been extensively studied. Kontio et al. and Bland independently synthesized the material, revealing a triclinic structure in their findings \cite{kontio1980, bland1958}.
Our theoretical investigation, using the PBEsol exchange-correlation (xc) functional, confirms that the energy-minimized structure adopts a triclinic (P$\overline{1}$) symmetry, in agreement with experimental findings. {The unit cell of \MAl~ is presented in Figure~\ref{fig:MnAl}(a).} The optimized lattice constants are $a = 4.991~\text{Å}$, $b = 8.672~\text{Å}$, $c = 4.986~\text{Å}$, and the angles are $\alpha = 90.24^\circ$, $\beta = 100.12^\circ$, and $\gamma = 105.15^\circ$. These values are in good agreement with the experimental parameters: $a = 5.092~\text{Å}$, $b = 8.862~\text{Å}$, $c = 5.047~\text{Å}$, and $\alpha = 85.19^\circ$, $\beta = 100.24^\circ$, $\gamma = 105.20^\circ$ \cite{pearson1958lattice, murray1987stable, PhysRevB.109.064207}.

{\par}In the triclinic \MAl~ structure, the inversion center is located at $(0, 0, 0)$, resulting in two inequivalent Mn atoms, labeled Mn$_1$ and Mn$_2$. The Al atoms occupy six distinct sites, denoted Al$_1$ to Al$_6$, with Al$_1$ to Al$_5$ having inversion-symmetric counterparts, while Al$_6$ resides at the inversion center. Both Mn$_1$ and Mn$_2$ are coordinated in a bicapped square antiprismatic environment, surrounded by ten Al atoms positioned at the vertices of a gyroelongated square pyramid. These ten Mn-Al bond lengths range from $2.37~\text{Å}$ to $2.75~\text{Å}$. Differences in coordination between Mn atoms are observed at the second nearest neighbor (NN) level. Mn$_1$ has two equidistant second nearest neighbors (Mn$_1$ and Mn$_2$) at $3.17~\text{Å}$, while Mn$_2$ has three second NN-one Mn$_1$ and two Mn$_2$-at distances of $3.17~\text{Å}$ and $3.26~\text{Å}$, respectively. A detailed summary of the coordination environments is provided in Table~\ref{tab:coordination_bonding}. Table~S2 of the \textit{supplement} tabulates the structural parameters in detail.

\subsubsection{Electronic and Magnetic Properties}
In its ground state, the \MAl~ intermetallic compound is found to be a half-metallic ferrimagnet. The magnetic moments of Mn$_1$ and Mn$_2$ are $-0.12~\mu_B$ and $+0.69~\mu_B$, respectively, while the Al atoms do not exhibit any significant magnetic behavior. As shown in Figure~\ref{fig:MnAl}(b), the down-spin channel ($\downarrow$) is metallic, while the up-spin channel ($\uparrow$) exhibits a gap of approximately $0.22~eV$ at E$_F$. The half-metallic nature of the material arises from a single band crossing at E$_F$ in the down-spin channel. 
This band is isolated from other bands in the CB region by an energy of $0.08~eV$ ($0.33~eV$ at Y to $0.41~eV$ at $\Gamma$ with respect to E$_F$). The contribution from this band appears as a peak in the DOS plot. There is an electron pocket along $\Gamma-$T$_2$ line near T$_2$. Given the two-fold symmetry of the system, there are two electron pockets in the full Brillouin Zone (BZ).
From the partial DOS (pDOS) plots of the 3$d$ orbitals of Mn$_1$ and Mn$_2$, it is observed that the isolated band near E$_F$ is primarily formed by the 3$d_{z^2}$ orbital of Mn$_2$. The sharp peak in the CB near E$F$ in the down-spin channel is mainly attributed to this orbital. From these pDOS plots in Figure~\ref{fig:MnAl}(d), it is concluded that the anisotropic behavior of the 3$d_{z^2}$ orbital of Mn$_2$ in both the up- and down-spin channels contributes significantly to the higher magnetic moment of Mn$_2$ compared to Mn$_1$.  
The orbital-projected band structure in Figure~S6 of the \textit{supplement} also confirms that the band at E$_F$ is primarily contributed by the Mn$_2$-3$d_{z^2}$ orbital. {Belin \etal has also reported that the Mn-3$d$ states cross the E$_F$ through X-ray emission spectroscopy technique \cite{belin2010}.} Interestingly, a mixed contribution from different orbitals to all the bands near E$_F$ is observed. The Al-3p orbital-projected band structure in Figure~S7 of the \textit{supplement}, over a wider energy range ($-6$ eV to $+6$ eV) around E$_F$, shows contribution, spread across the dense collection of bands. It suggests hybridization between the Mn-3$d$ and Al-3$p$ states. Further bonding analysis, discussed later, will provide more insight into this.

{\par} Half-metallic ferrimagnetism, which is typically observed in disordered systems and Heusler alloys, is relatively rare in intermetallics \cite{ozdog2007,ostler2011}. The occurrence of half-metallic ferrimagnetism in these alloys is often linked with the SOC, which can play a crucial role in controlling the switching of magnetic states \cite{je2018}. However, other interactions like superexchange or Ruderman-Kittel-Kasuya-Yosida (RKKY) play role in several such systems. In the case of \MAl, our DFT calculations, including SOC effects, confirm that neither the Mn nor Al atoms show a significant SOC effect (see Figure~S1 of \textit{supplement}). Further verification of the system through additional tests indicates no involvement of non-collinear magnetic characters in this ferrimagnetic system.  A comparison of the electronic and magnetic characteristics of \MAl~ with various Mn-Al based alloys is presented in Table~\ref{tab:comparison}. Among these, a ferrimagnetic half-metallic character is observed in Mn$_2$VAl and Mn$_2$CrAl, with Mn$_2$CrAl exhibiting a similar magnetic moment and gap \cite{jiang2001,jum2019}.

{\par} The electronic structure analysis reveals that the magnetic properties near the E$_F$ are primarily governed by the Mn atoms. The total DOS (TDOS) in the valence region shows peaks at approximately $-0.6~eV$ for the up-spin channel and $-1.2~eV$ for the down-spin channel. The pDOS of Al atoms remains almost symmetrical for both spin channels, which indicates their non-magnetic nature. Furthermore, the absence of contributions from Al atoms near E$_F$ aligns with the experimental observation of an Al pseudo-gap in \MAl \cite{belin2010}. The 3$d$ orbitals of Mn are more localized compared to the 3$p$ orbitals of Al, with the Mn pDOS rapidly decreasing as we move away from the Fermi level, while the Al pDOS spreads more evenly.

{\par} Further analysis reveals that Mn$_1$ does not contribute significantly to the half-metallic behavior of the system. Both Mn$_1$ and Mn$_2$ exhibit distinct features in their $t{_2g}$ orbitals, specifically in the 3$d_{xy}$, 3$d_{zx}$, and 3$d_{zy}$ orbitals, which display different pDOS profiles for both spin channels. This behavior is also observed in the $e_g$ orbitals, indicating a Jahn-Teller effect, which is expected in a system with such low crystal symmetry. Notably, the sharp peak in the DOS near E$_F$ for the spin-up channel in Figure~\ref{fig:MnAl}(b) is mainly attributed to the 3$d_{xy}$ orbital of Mn$_1$, while the $e_g$ orbitals contribute different pDOS patterns for both Mn sites.

\begin{figure}[]
\centering	
\includegraphics[width=0.8\textwidth]{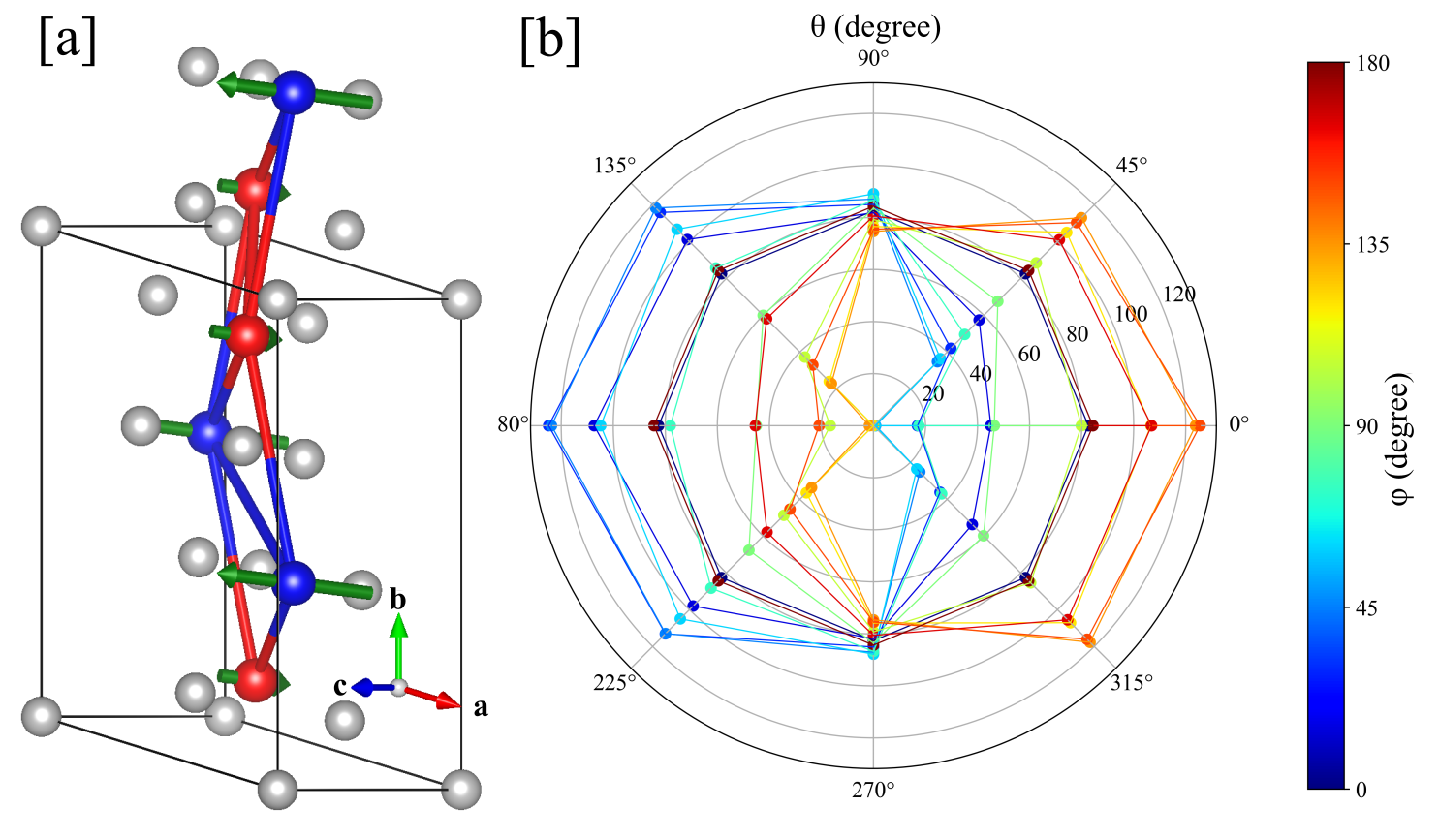}
\caption{\label{fig:MAE-Mag} (a) Magnetic configuration with lowest energy including spin--orbital coupling of \MAl; (b) Magneto-crystalline anisotropic energy variation with respect spherical polar coordinates $\theta$ and $\phi$ of \MAl.}
\end{figure}

\subsubsection{Magneto-crystalline Anisotropy }
{\par} The anisotropy is intrinsic to the crystal structure of \MAl, which leads to significant magneto-crystalline anisotropy (MAE), as shown in the plot in Figure~\ref{fig:MAE-Mag}(b). The maximum MAE of $125~\mu{eV}$ is observed at an angle of $\theta = 150^\circ$ and $\phi = 0^\circ$, with the magnetic easy axis aligned at $\theta = 120^\circ$ and $\phi = 180^\circ$. A 3-dimensional plot of MAE can be found in Figure~S1(c) of \textit{supplement}. This anisotropic behavior can be attributed to the structural arrangement of the Mn atoms, which form a slightly buckled layer, as seen in Figure~\ref{fig:MAE-Mag}(a). The buckling results in a small deviation from a perfectly planar structure, and the magnetic easy axis is found to be nearly perpendicular to this buckled layer. This strong correlation between the structural configuration of the Mn atoms and the magnetic anisotropy direction further underscores the critical role of crystal symmetry in determining the magnetic properties of \MAl.

\subsubsection{Effect of Strain}
\begin{figure}[]
\centering	
\includegraphics[width=\textwidth]{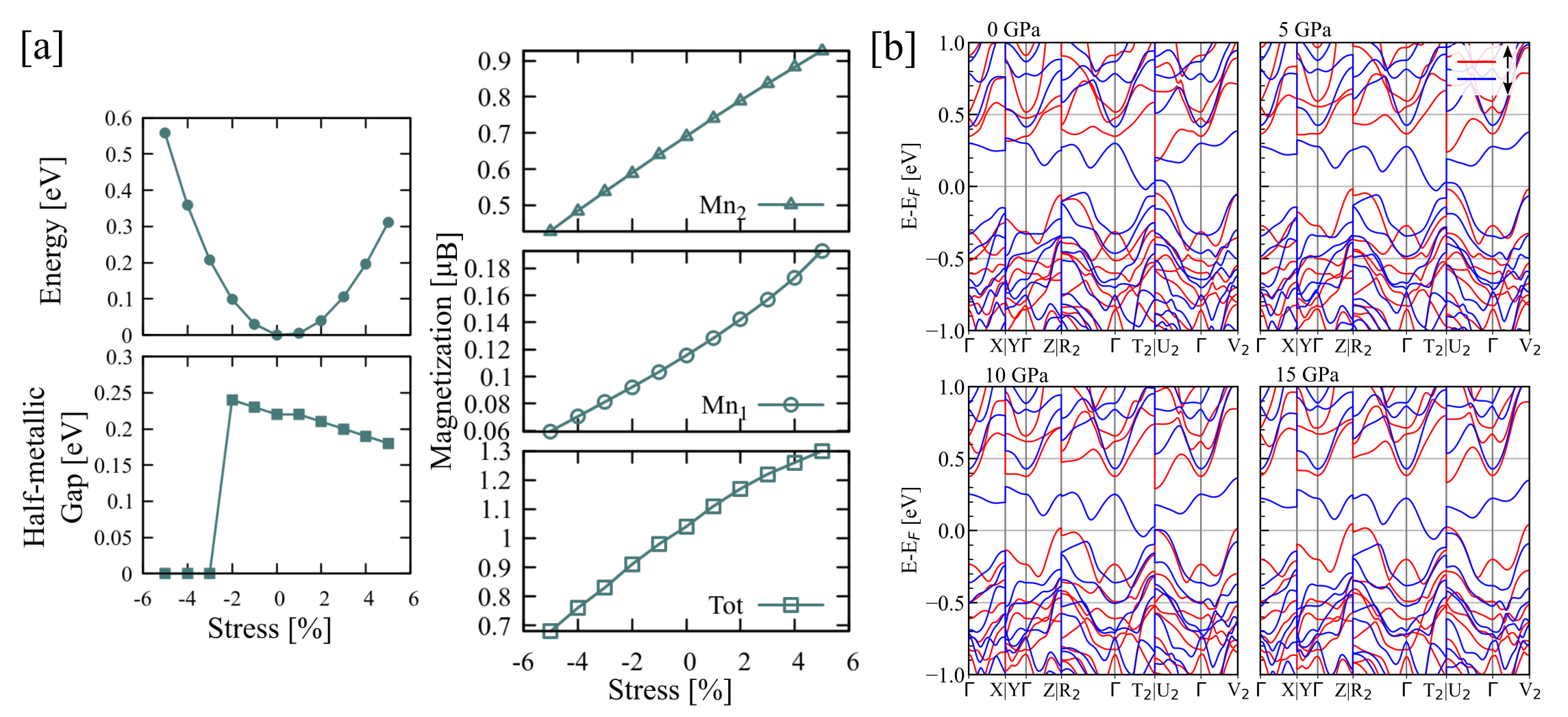}
\caption{\label{fig:stress_var}   (a) Variation of relative energy, half-metallic band-gap, absolute value of magnetization of Mn$_1$ $\&$ Mn$_2$ and the total magnetization per unit cell with uniaxial percentage strain along crystal axis $b$; (b) Band structure of \MAl~ under isotropic pressure.}	
\end{figure}

{\par}{The structure exhibits low symmetry; however, it appears to be layered along the $b$-axis. In layered structures, it is easier to apply strain along the direction perpendicular to the layering plane without significantly disturbing the internal coordination of atoms. This is why the b-axis was chosen-offering a simple and straightforward way to probe the robustness of the system’s electronic and magnetic behavior. Also, the Mn-Mn interaction within a layer is much stronger than between the layers. This was also experimentally observed by Dunlop \etal \cite{dunlop1976}.}
{\par} The electronic and magnetic properties of \MAl~ are examined under uniaxial strain along the $b$-axis. A compression of over $2\%$ ($-2\%$ strain) induces a transition from half-metallic to metallic behavior. A detailed analysis of the band dispersion for the strained structures (Figure~S2) can be found in \textit{Section III} of the \textit{supplement}. The magnetization of both Mn$_1$ and Mn$_2$, as well as the total magnetization, exhibits a steady parabolic variation, as shown in Figure~\ref{fig:stress_var}(a). Interestingly, the trends for Mn$_1$ and Mn$_2$ are opposite, with the total magnetization following the behavior of Mn$_2$. In contrast to the total energy variation, magnetization reaches its lowest value at $-5\%$ strain and its highest at $+5\%$ strain. So, with uniaxial strain, we cannot induce any magnetic transition; only a variation of magnetic moments can be achieved for \MAl. In contrast, a half-metal to metal transition can be achieved with minimal compression. This suggests that the ferrimagnetic character of \MAl~ is more resilient than its half-metallic nature when subjected to uniaxial strain. {The strain perpendicular to the layers forces the Mn atoms in different layers interact more strongly, hence the half-metal to metal transition could be achieved.}

{\par}{An isotropic compression can also induce half-metal to metal transition but only over $10~ GPa$ pressure as seen in Figure~\ref{fig:stress_var}(b). At $10~ GPa$ the down-spin band at Z point crosses E$_F$. At higher pressure, band crossing is observed at U$_2$, V$_2$, and along the R$_2$-$\Gamma$ line for down-spin channel. The ferrimagnetic character remains intact even under $15~ GPa$. The magnetisation drops with pressure, from $1.04~\mu_B$ at $0~GPa$ to $0.71~ \mu_B$ at $15~ GPa$ (see Table~S1 of \textit{supplement}).} 

\begin{table*}[]
	\centering
	\setlength{\tabcolsep}{3pt}
	\caption{\label{tab:coordination_bonding} Structural parameters including lattice constants, angles, volume; bonding character including nearest neighbor information (NNB); electronic properties including the density of states/cell at E$_F$ for both spin channels within bracket; and magnetization  of \MAl, \MGe, and \MG~ structures.}
	\begin{tabular}{cccccccc} 
		\hline\hline
		\textbf{System} & \begin{tabular}[c]{@{}c@{}}a,b,c ($\AA$) \\ $\alpha$,$\beta$,$\gamma$ ($^o$) \\ V ($\AA^3$)\end{tabular} & \textbf{Bonds}  & \textbf{NNB} & \begin{tabular}[c]{@{}c@{}}\textbf{Antibond} \\ \textbf{below } $E_F$\end{tabular}  & \begin{tabular}[c]{@{}c@{}}\textbf{Electronic} \\ \textbf{Character}\end{tabular} & \begin{tabular}[c]{@{}c@{}}\textbf{Magn.}  \\ \textbf{cell, Mn$_1$, Mn$_2$} \\ \textbf{($\mu_B$)}\end{tabular} \\ 
		\hline
		\multirow{3}{*}{Mn$_4$Al$_{11}$} & \multirow{3}{*}{\begin{tabular}[c]{@{}c@{}}4.991, 8.672, 4.985 \\ 90.24, 100.12, 105.15 \\ 204.76\end{tabular}} & Mn$_1$-Al& 10 & 1.201\%  & \multirow{2}{*}{\begin{tabular}[c]{@{}c@{}}Half-Metal \\ (0.00, 1.74)\end{tabular}} & \multirow{2}{*}{1.04, -0.12, 0.69} \\
		& & Mn$_2$-Al & 10 & 0.0\% & & & \\
		& &  &  &  & & & \\
		\hline
		
		\multirow{4}{*}{Mn$_4$Al$_9$Ge$_2$-1} & \multirow{4}{*}{\begin{tabular}[c]{@{}c@{}}5.013, 8.708, 4.950 \\ 90.42, 100.03, 105.06 \\ 205.15\end{tabular}} & Mn$_1$-Al & 7 & 2.542\%  & \multirow{4}{*}{\begin{tabular}[c]{@{}c@{}} Metal \\ (3.87, 2.61)\end{tabular}} & \multirow{4}{*}{0.22, -0.02, 0.13} \\
		& & Mn$_1$-Ge & 3 & 3.030\% & & & \\
		& & Mn$_2$-Al & 9 & 1.141\% & & & \\
		& & Mn$_2$-Ge & 1 & 2.857\% & & & \\
		\hline
		
		\multirow{4}{*}{Mn$_4$Al$_9$Ge$_2$-2} & \multirow{4}{*}{\begin{tabular}[c]{@{}c@{}}4.920, 8.675, 5.047 \\ 89.94, 99.71, 104.91 \\ 204.98\end{tabular}} & \multirow{4}{*}{\begin{tabular}[c]{@{}c@{}}Mn$_1$-Al \\ Mn$_1$-Ge \\ Mn$_2$-Al \\ Mn$_2$-Ge\end{tabular}} & 7 & 3.017\% &  \multirow{4}{*}{\begin{tabular}[c]{@{}c@{}} Metal \\ (5.72, 2.10)\end{tabular}} & \multirow{4}{*}{0.42, 0.02, 0.21} \\
		& & & 3 & 4.854\% & & & \\
		& & & 9 & 5.694\% & & & \\
		& & & 1 & 3.571\% & & & \\
		\hline
		
		\multirow{4}{*}{Mn$_4$Al$_9$Ge$_2$-3} & \multirow{4}{*}{\begin{tabular}[c]{@{}c@{}}4.950, 8.701, 5.005 \\ 90.09, 100.30, 104.67 \\ 204.96\end{tabular}} & \multirow{4}{*}{\begin{tabular}[c]{@{}c@{}}Mn$_1$-Al \\ Mn$_1$-Ge \\ Mn$_2$-Al \\ Mn$_2$-Ge\end{tabular}} & 9 & 2.667\% &  \multirow{4}{*}{\begin{tabular}[c]{@{}c@{}} Metal \\ (3.94, 3.36)\end{tabular}} & \multirow{4}{*}{0.08, 0.00, 0.04} \\
		& & & 1 & 2.778\% & & & \\
		& & & 7 & 1.442\% & & & \\
		& & & 3 & 3.297\% & & & \\
		\hline
		
		\multirow{4}{*}{Mn$_4$Al$_9$Ge$_2$-4} & \multirow{4}{*}{\begin{tabular}[c]{@{}c@{}}5.026, 8.716, 4.996 \\ 90.36, 101.00, 106.90 \\ 205.11\end{tabular}} & \multirow{4}{*}{\begin{tabular}[c]{@{}c@{}}Mn$_1$-Al \\ Mn$_1$-Ge \\ Mn$_2$-Al \\ Mn$_2$-Ge\end{tabular}} & 9 & 2.055\%  & \multirow{4}{*}{\begin{tabular}[c]{@{}c@{}} Metal \\ (4.25, 1.81)\end{tabular}} & \multirow{4}{*}{0.49, 0.02, 0.24} \\
		& & & 1 & 5.128\% & & & \\
		& & & 7 & 1.042\% & & & \\
		& & & 3 & 5.769\% & & & \\
		\hline
		
		\multirow{4}{*}{Mn$_4$Al$_9$Ge$_2$-5} & \multirow{4}{*}{\begin{tabular}[c]{@{}c@{}}5.015, 8.668, 4.997 \\ 89.45, 102.06, 104.46 \\ 205.50\end{tabular}} & \multirow{4}{*}{\begin{tabular}[c]{@{}c@{}}Mn$_1$-Al \\ Mn$_1$-Ge \\ Mn$_2$-Al \\ Mn$_2$-Ge\end{tabular}} & 9 & 3.390\%  & \multirow{4}{*}{\begin{tabular}[c]{@{}c@{}} Metal \\ (4.71, 0.97)\end{tabular}} & \multirow{4}{*}{0.70, 0.06, 0.30} \\
		& & & 1 & 4.878\% & & & \\
		& & & 8 & 0.450\% & & & \\
		& & & 2 & 2.667\% & & & \\
		\hline
		
		\multirow{3}{*}{Mn$_4$Al$_{10}$Ge} & \multirow{3}{*}{\begin{tabular}[c]{@{}c@{}}4.993, 8.699, 5.013 \\ 90.70, 101.24, 105.54 \\ 205.30\end{tabular}} & \multirow{3}{*}{\begin{tabular}[c]{@{}c@{}}Mn$_1$-Al \\ Mn$_1$-Ge \\ Mn$_2$-Al\end{tabular}} & 9 & 2.182\%  & \multirow{3}{*}{\begin{tabular}[c]{@{}c@{}} Semiconductor \\ (0.00, 0.00)\end{tabular}} & \multirow{3}{*}{0.07, -0.02, 0.05} \\
		& & & 1 & 4.615\% & & & \\
		& & & 10 & 2.606\% & & & \\
		\hline
		\hline
	\end{tabular}
\end{table*}

\begin{table*}[]
	\caption{\label{tab:comparison} Comparison of magnetic and electronic property of \MAl~ with different alloys \cite{jamer2017,jiang2001,jum2019,luo2008,ouardi2013,kourov2015,ishida1995,ze2017,deb2000}.}
	\begin{tabular}{cccccc}
		\hline \hline
		Compounds & \begin{tabular}[c]{@{}c@{}}Magnetic\\ Character\end{tabular} & \begin{tabular}[c]{@{}c@{}}Magnetic\\  Moment ($\mu_B$)\end{tabular} & \begin{tabular}[c]{@{}c@{}}Electronic\\ Character\end{tabular} & \begin{tabular}[c]{@{}c@{}}Electronic\\ gap (eV)\end{tabular} & Reference \\ \hline
		Mn$_4$Al$_{11}$ & Ferrimagnet & 1.04 & half-metallic & 0.22 & - \\
		Mn$_3$Al & \begin{tabular}[c]{@{}c@{}}Compensated\\ Ferrimagnet\end{tabular} & 0.00 & half-metallic & 0.40 & \cite{jamer2017} \\
		Mn$_2$VAl & Ferrimagnet & 1.94 & half-metallic & 0.13 & \cite{jiang2001} \\
		Mn$_2$CrAl & Ferrimagnet & 1.03 & half-metallic & 0.21 & \cite{jum2019} \\
		Mn$_2$FeAl & Ferrimagnet & 1.01 & metallic & - & \cite{luo2008} \\
		Mn$_2$CoAl & Ferrimagnet & 2.00 & \begin{tabular}[c]{@{}c@{}}spin-gapless\\ semiconductor\end{tabular} & - & \cite{ouardi2013} \\ 
		Co$_2$MnAl & Ferromagnetic & 4.17 & half-metallic & 0.66 & \cite{kourov2015,ishida1995} \\
		Fe$_2$MnAl & Ferromagnetic & 1.32 & half-metallic & 0.55 & \cite{kourov2015, ze2017} \\
		Cu$_2$MnAl & Ferromagnetic & 4.12 & metal & - & \cite{deb2000} \\ \hline \hline
	\end{tabular}
\end{table*}

\subsection{Effect of Ge doping in Mn$_{4}$Al$_{11}$}
{We have chosen Ge to substitute Al in \MAl~ as Ge introduces an additional valence electron and has a larger atomic size compared, both of which could influence the Mn-Al/Ge bonding interactions.
Notably, the site-selective substitution of Ge will allow us to assess role of electronic-structure change on thermodynamic stability and structural properties due to chemically altered Mn-environment in \MAl.
We have also checked the formation energy of Si-doped systems and found that Ge-substituted structures are thermodynamically favorable (see Figure~S3 of \textit{supplement}).
The anomalous looking formation energy trend in Ge-doped \MAl~ is traced to arise from a combination of:  (a) site-specific strain effects (due to larger atomic radius of Ge ($125~pm$) compared to Si ($111~pm$), and (b) orbital sensitivity to site symmetry (leading to stronger Mn-Ge hybridization), which are not as pronounced in Si due to its smaller size and weaker hybridization.}   

\subsubsection{Structure}
When doping with Ge atoms at the Al sites in \MAl, a total of six distinct structures are observed, labeled \MGe-1 to \MGe-5, depending on which Al site (Al$_1$ to Al$_5$) is replaced by Ge, and the \MG~ structure, where Al$_6$ is replaced by Ge. {For \MGe~ the substitutional doping concentration is $2/11\simeq 18.18\%$ and for  \MG~ the doping concentration is $1/11 \simeq 9.09\%$.} In all Ge-doped \MAl~ structures, the overall volume of the systems increases due to the larger atomic radius of Ge compared to Al. However, the overall shape of the structures does not undergo significant distortion, as evidenced by the lattice parameters presented in Table~\ref{tab:coordination_bonding}. The increased size of the Ge atom leads to noticeable distortions in the local atomic arrangement, which are reflected in the bond lengths, as discussed in Table~S2 of the \textit{supplement}. Specifically, the Mn-Al bond lengths increase upon Ge doping, while the Mn-Ge bond lengths in the doped systems are shorter than the Mn-Al bond lengths in the pristine \MAl~ structure. This behavior is consistent with the larger atomic radius of Ge, which compresses the space between Mn atoms when it replaces Al, influencing the bonding environment in the alloy.

{\par} When a Ge atom replaces the Al$_6$ atom at the inversion center ($000$), the resulting structure is labeled as \MG. This doping site is particularly significant because it is positioned centrally between the Mn$_1$ and Mn$_2$ atoms, which plays a crucial role in influencing the magnetic and electronic properties of the alloy. The lattice constants and angles of the \MG~ structure remain almost identical to those of the pristine \MAl~ structure, indicating minimal distortion in the overall crystal symmetry.The variation in bond lengths follows a trend similar to that observed in other doped systems.

\begin{figure}[h!]
	\centering	
	\includegraphics[width=0.7\textwidth]{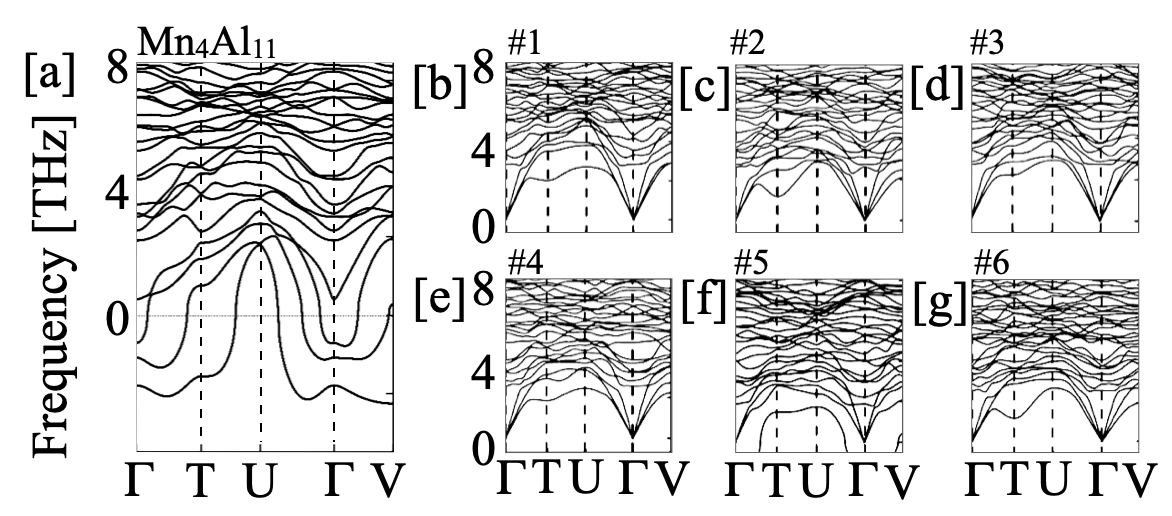}
	\caption{\label{fig:phonons}  {The phononic band diagram of (a) \MAl, (b-f) \MGe-1 to \MGe-5, and (g) \MG.}}	
\end{figure}

{\par} {The dynamical stability of all the structures has been verified through phonon dispersion analysis. Our phonon calculations of \MAl in Figure~\ref{fig:phonons}(a) shows dynamical instability as reflected through imaginary modes. Interestingly, though \MAl has been successfully synthesized, it is mostly found in mixed phase or even in a metastable phase \cite{ khvan2015formation, xian2021al11mn4}. This means it can form under certain conditions such as rapid quenching, where atoms do not have enough time or mobility to rearrange into a more stable structure. In Figure~\ref{fig:phonons}(b-g), we show that site-selective alloying by Ge at different inequivalent Al sites (\MGe-1 to \MGe-4, and \MG) eliminates imaginary phonon modes, indicating enhanced lattice stability. The Ge alloying suppresses unstable vibrational modes by altering local Mn-Al/Mn-Ge bonding strength, and electronic structure. Among the Ge doped systems, only \MGe-5 exhibits negative phonon modes. The reason for the dynamical instability of \MGe-5 is discussed later using a bonding-antibonding analysis.}

\subsubsection{Electronic and Magnetic Properties}
\begin{figure*}[]
	\centering	
	\includegraphics[width=\textwidth]{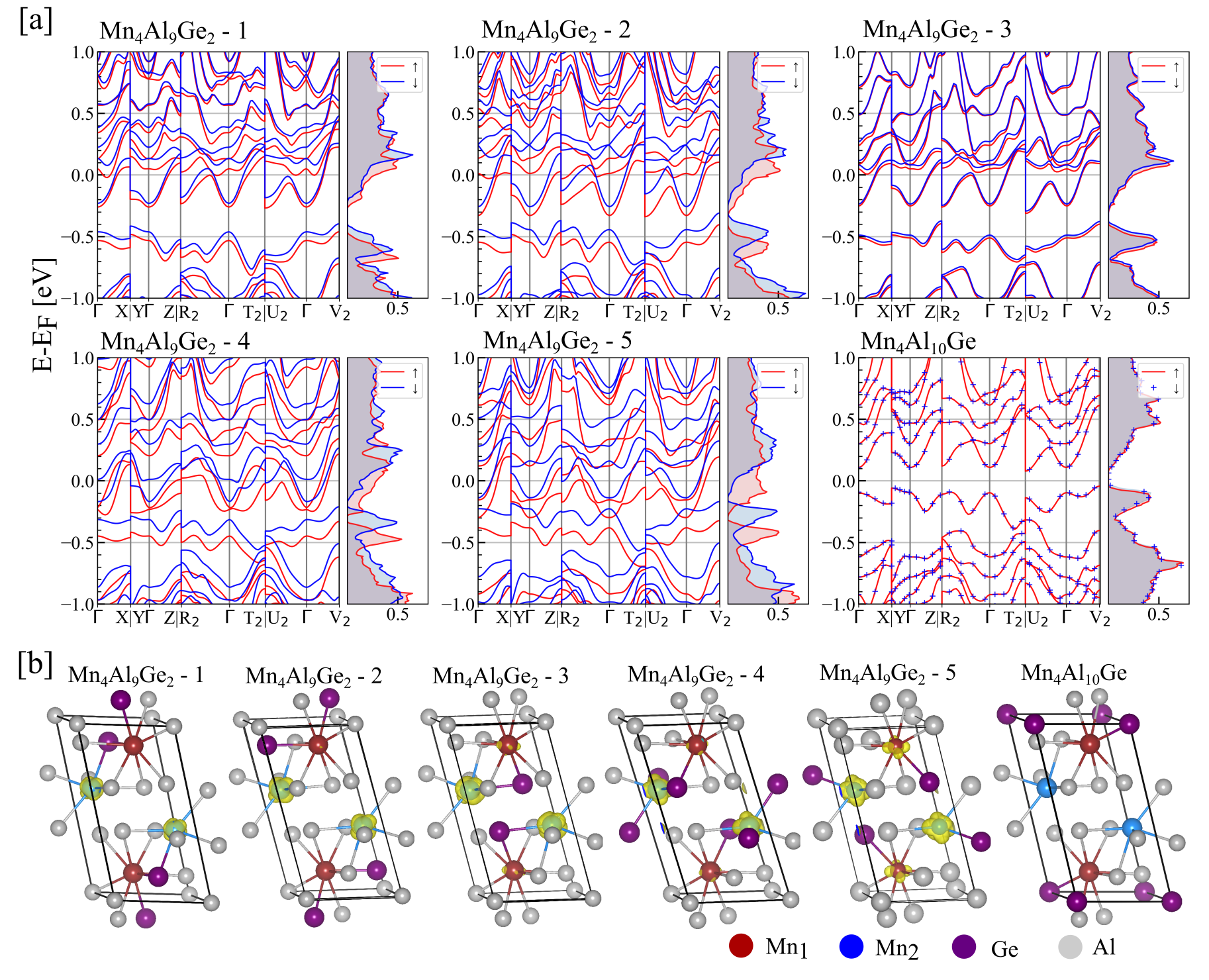}
	\caption{\label{fig:MnAlGe1to6-band} (a) Band structure and DOS of different Ge doped \MAl~ when Ge replaces Al$_1$ to Al$_6$ sites, identified as Mn$_{4}$Al$_{9}$Ge$_2$-1 to Mn$_{4}$Al$_{9}$Ge$_2$-5, and Mn$_{4}$Al$_{10}$Ge; (b) The difference of up and down electron densities ($\rho_\uparrow-\rho_\downarrow$) of these Ge doped structures.}	
\end{figure*}

{\par} The electronic band dispersion and DOS for the Ge-doped \MAl~ alloys (\MGe-1 to \MGe-5) are shown in Figure~\ref{fig:MnAlGe1to6-band}, revealing that all the doped structures exhibit a finite DOS at E$_F$, indicating a transition to metallic behavior. This confirms that Ge doping effectively disrupts the half-metallic character observed in the pristine \MAl~ structure. In these doped systems, a gap is observed just below E$_F$, signifying a band separation in the higher valence band region. The band responsible for the half-metallic nature in \MAl, which is isolated at E$_F$ in the undoped structure, is no longer present as a separate band in the Ge-doped systems. The band dispersions in the doped alloys are also sharper, suggesting increased electron localization.

{\par}{For quantitative assessment of the metallicity DOS/cell at E$_F$, N(E$_F$) are calculated for both pristine and doped structures. Clearly, a higher N(E$_F$) value is indicative of enhanced metallic character, while a near-zero or vanishing N(E$_F$) correlates with semiconducting or insulating behavior. Looking at the values of N(E$_F$) presented in Table~\ref{tab:coordination_bonding}, we can clearly confirm that the vanishing magnetic character in \MGe-3 is directly correlated with the diminishing difference in N(E$_F$) between the two spin channels. As the difference in N(E$_F$) between the up and down spins increases, the magnetization also increases.}

{\par} In particular, for \MGe-1, \MGe-2, \MGe-4, and \MGe-5, the up and down spin bands exhibit distinct characteristics, with their projected DOS showing a clear separation near E$_F$. However, as we move away from E$_F$, the distinction between the up- and down-spin channels gradually diminishes. This behavior highlights the weakening of the spin polarization with increasing energy. In contrast, the \MGe-3 structure displays nearly degenerate spin-up and spin-down bands, with their respective projected DOS almost overlapping, suggesting a more non-magnetic character for this configuration.
Thus, while Ge doping disrupts the half-metallic behavior, it introduces complex variations in the electronic properties depending on the doping site.

{\par} {The pDOS shown in Figure~S5 of the \textit{supplement} reveal that Mn atoms (particularly, partially filled Mn-3$d$ orbitals) contribute significantly near E$_F$ in all the Ge-doped \MAl~ structures, while Ge itself does not contribute to the electronic states near E$_F$ (see Figure~S4 of the \textit{supplement}).} In particular, the 3$d_{z^2}$ band of Mn$_2$ remains more localized than other Mn 3$d$ orbitals across all the \MGe ~systems, a feature that mirrors the behavior observed in the pristine Mn$_3$Al structure. The key difference, however, is that in the doped systems, this localized 3$d_{z^2}$ band shifts into the valence band (VB) region, whereas in the undoped system, it is positioned closer to the Fermi level. This shift suggests that Ge doping induces a modification in the electronic structure, affecting the localization of Mn's 3$d_{z^2}$ orbital and altering the overall band structure in the process.

{\par} Except for \MGe-5, none of the Ge-doped \MAl~ structures exhibit as strong a magnetic character as the pristine \MAl~ crystal. In all cases, the magnetic nature remains confined to the Mn atoms, with Mn$_2$ carrying a higher magnetic moment than Mn$_1$, similar to the undoped structure. The electron density difference plots shown in Figure~\ref{fig:MnAlGe1to6-band}(b) provide further evidence of the role of Mn$_2$ in carrying the magnetic character. These plots highlight a significant electron density difference for the \MGe-2, \MGe-4, and \MGe-5 systems, which corresponds with the asymmetry in the up and down spin DOS observed for these structures. This confirms that the magnetic behavior is primarily influenced by Mn$_2$, and the doping with Ge affects the symmetry and strength of the magnetic properties in a manner consistent with the observed electronic structure changes. {\MGe-5 which shows highest magnetization/cell of value $0.70~\mu_B$ also exhibits a high value of MAE of $525~\mu eV$ for $\theta=90^o$ and $\phi=45^o$  (see, Figure~S12 of the \textit{supplement}).}

{\par} The electronic-structure of \MG~ is depicted through band structure and DOS plots in Figure~\ref{fig:MnAlGe1to6-band}(a). In contrast to the pristine \MAl, the substitutional doping by Ge in \MG~ eliminates ferrimagnetic order at the same time induces MIT. This is also reflected through overlapping band-structure and DOS in both up/down-spin channels resulting from reduced magnetic-exchange coupling to zero due to Ge substitution at Al$=(000)$. 
The excess electron from Ge leads to filling of partially occupied Mn$-d$ bands near the Fermi-level, which is responsible for the half-metallicity in \MAl~ now resides just below E$_F$ in \MG, creating an almost distinct hump in the DOS. 
Similar to the pristine \MAl structure, the pDOS in Figure~S5 of \textit{supplement} show that the 3$d_{z^2}$ orbital of Mn$_2$ is still the main contributor. 
However, the pDOS is symmetrical for both up and down spin channels, signaling the loss of magnetic polarization. The difference in the up- and down-spin electron density plot in Figure~\ref{fig:MnAlGe1to6-band}(b) supports this observation. Moreover, the nature of the pDOS for Mn$_1$ and Mn$_2$ is drastically different, with the 3$d$ orbitals of Mn$_2$ being more localized than those of Mn$_1$, and the former contributing predominantly near the Fermi level. The calculated band-gap is  $0.14~eV$, which is indirect in nature (Y $\rightarrow$ U$_2$).
{The sharpness in band dispersions in \MG~ relative to \MAl~ arises from enhanced Mn-Ge orbital hybridization, as will also be discussed with the help of COHP results. This is primarily attributed to the higher electronegativity of Ge and its more delocalized 4$p$ orbitals compared to Al. The more coherent bonding interactions in \MG, along with reduced electronic scattering, result in cleaner and more dispersive bands near E$_F$.}

\begin{figure}[]
	\centering	
	\includegraphics[width=0.7\textwidth]{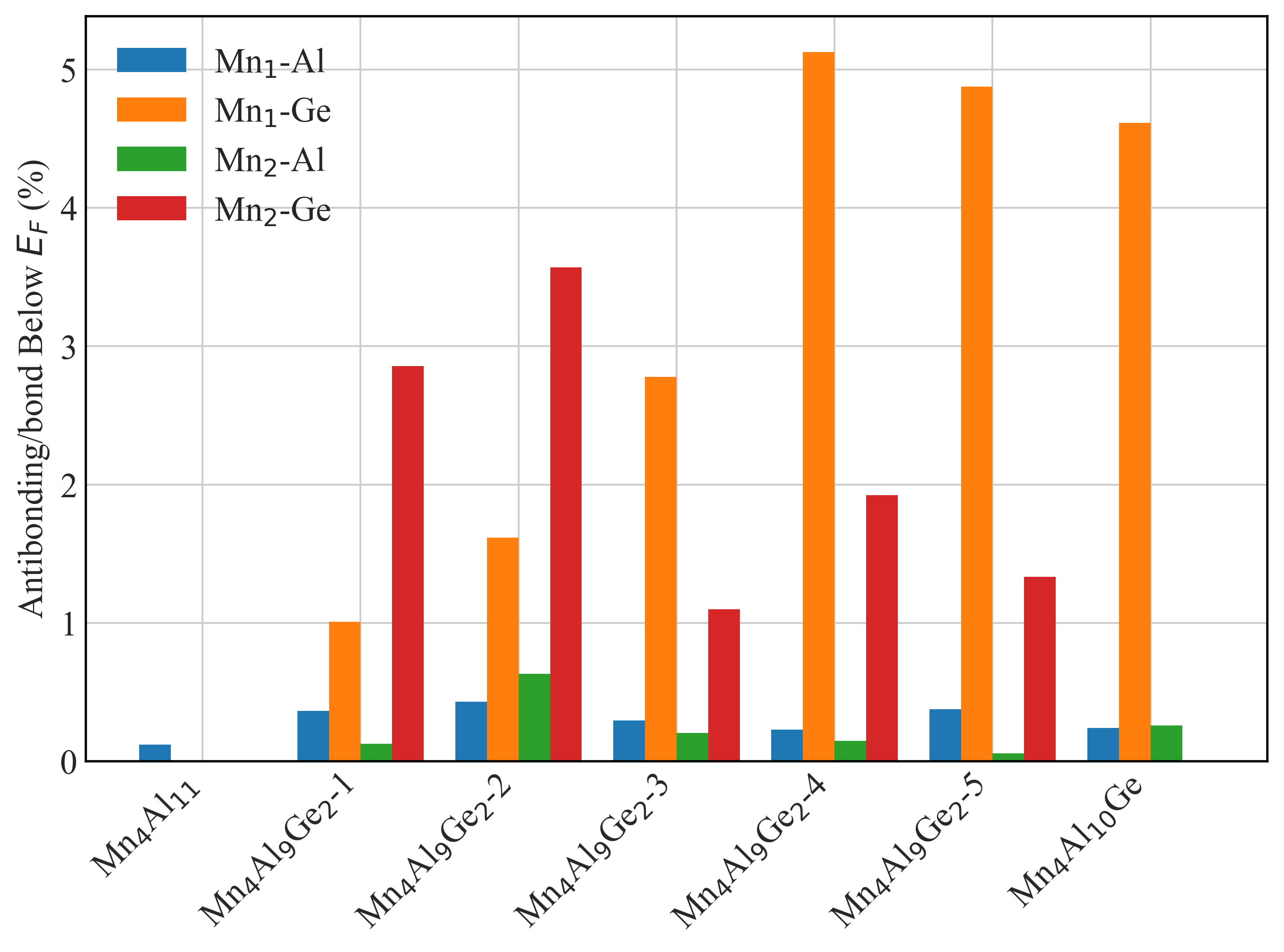}
	\caption{\label{fig:antibonding} Percentage of antibonding per bond for \MAl, \MGe, and \MG~ below Fermi level.}	
\end{figure}

\begin{table}[]
	\centering
	\resizebox{\textwidth}{!}{
	\begin{tabular}{c|cccccccccc}
		\hline \hline
		\textbf{System} & \textbf{Mn$_1$} & \textbf{Mn$_2$} & \textbf{Al$_1$} & \textbf{Al$_2$} & \textbf{Al$_3$} & \textbf{Al$_4$} & \textbf{Al$_5$} & \textbf{Al$_6$} & \textbf{Ge} \\ \hline
		\textbf{Mn$_4$Al$_{11}$} & -3.49 & -2.33 & 1.14 & 0.98 & 1.01 & 0.98 & 1.09 & 1.22 & --\\
		\textbf{Mn$_4$Al$_9$Ge$_2$-1} & -2.20 & -1.93 & -- & 1.03 & 0.99 & 1.14 & 0.99 & 1.22 & -0.51\\
		\textbf{Mn$_4$Al$_9$Ge$_2$-2} & -2.14 & -2.27 & 1.17 & -- & 1.19 & 1.08 & 0.98 & 1.18 & -0.61\\
		\textbf{Mn$_4$Al$_9$Ge$_2$-3} & -2.97 & -1.34 & 1.08 & 1.13 & -- & 1.15 & 1.29 & 1.12 & -0.90\\
		\textbf{Mn$_4$Al$_9$Ge$_2$-4} & -2.77 & -1.21 & 1.21 & 1.03 & 0.88 & -- & 1.24 & 1.20 & -0.99\\
		\textbf{Mn$_4$Al$_9$Ge$_2$-5} & -2.71 & -1.32 & 1.08 & 0.99 & 1.14 & 1.14 & -- & 1.29 & -0.96\\		
		\textbf{Mn$_4$Al$_{10}$Ge} & -2.43 & -2.43 & 1.15 & 0.99 & 0.98 & 1.03 & 1.08 & --& -0.74 \\
		\hline \hline 
	\end{tabular}}
	\caption{ \label{tab:bader} Bader charges of the atoms for \MAl, \MGe, and \MG.}
\end{table}

\noindent

\section{Discussion}
The local chemical bonding plays a crucial role in understanding the nature of bonding, anti-bonding, and non-bonding interactions within materials \cite{reitz2024bonding,tolborg2021chemical}. In the context of localized basis sets, the calculation of atomic orbital overlap can be easily achieved using the crystal orbital overlap population (COOP) method \cite{dronskowski2004}. This technique provides insights into the strength and character of the interaction between atomic orbitals in a material. For DFT calculations utilizing plane wave basis sets, the COHP method is employed to assess the bonding interactions.
{Energy integral of the COHP (ICOHP) quantifies the contribution of a specific bond to the band energy. The ICOHP values were calculated by integrating spin-up and spin-down contributions of COHP separately. Here, ICOHP reflects the nature of bonding or hybridization among the alloying species, where a higher value indicates a more covalent character. In other words, the ICOHP till the E$_F$ is a powerful indicator of bond strength.} 

{\par} Table~\ref{tab:coordination_bonding} provides a comprehensive overview of the coordination environments and bonding interactions in the Mn$_4$Al$_{11-x}$Ge$_x$ compounds. This data highlights the diverse nature of bonding interactions, reflecting variations in antibonding interactions and coordination environments across different systems.
As shown in Figure~\ref{fig:antibonding}, the antibonding percentage per bond below the E$_F$ for \MAl~ is quite low compared to the doped systems. Referring to Figure~S8 in the \textit{supplement}, we observe that, in addition to the first NN Mn-Al bonding, the second NN Mn-Mn bonding character is also significant. Table~S2 in the \textit{supplement} shows that the ICOHP values are small, ranging from $-0.43$ to $-0.88$. This suggests a less covalent and more metallic nature of the bonding. Additionally, Figure~S9 in the \textit{supplement} reveals that the Mn$_1$-Al$_6$ bonding-antibonding interaction is the strongest, which is also supported by the highest ICOHP value of $-0.88$ for \MAl. The COHP analysis shows that Mn$_1$-Al interactions near the Fermi level (ranging from $-0.4~eV$ to $0.4~eV$ with respect to E$_F$) remain nearly equivalent in both spin-up and spin-down states, resulting in an absence of significant spin asymmetry in the electronic structure. This suggests that Mn$_1$ contribution to the DOS near E$_F$ is largely non-polarized, confirms that it does not play a major role in the formation of the half-metallic gap. In contrast, Mn$_2$ exhibits strong spin-dependent hybridization with Al. The COHP results demonstrate that in the spin-up channel, Mn$_2$-Al interactions show predominantly bonding character, while in the spin-down channel, the interactions become strongly anti-bonding. This asymmetry is a key factor in opening the half-metallic gap, as it suppresses electronic states at the Fermi level in one spin channel while allowing metallic behavior in the other.

{\par}In doped systems, Al-Ge interactions are notably elevated, peaking in \MGe-4. In the structures from \MGe-3 to \MGe-5, Mn$_1$ forms a singular Mn-Ge bond, while Mn$_2$ engages in three Mn-Ge bonds, resulting in a significantly higher antibonding state per bond below E$_F$ for Mn$_1$-Ge compared to Mn$_2$-Ge. Further analysis in Figure~S8 of \textit{supplement} reveals distinct differences in the COHP profiles for these bonds, indicating that Ge doping induces localized changes in the bonding characteristics, which in turn, significantly influences the electronic and magnetic properties of the compounds. The Mn-Mn COHP profiles for all compounds, except \MG, demonstrate the importance of second NN interactions in shaping these properties. In \MG, however, the direct connection between Mn$_1$ and Mn$_2$ via the Ge atom facilitates a dominant superexchange mechanism, making the Mn$_1$-Mn$_2$ COHP effectively negligible (see, Figure~S10 and Figure~S11 of the \textit{supplement}).
Also, for \MG~ the Mn$_1$-Ge bonding-antibonding strength is strongest among all Mn$_{4}$Al$_{11-x}$Ge$_x$ compounds in discussion, as the ICOHP at E$_F$ is highest ($-1.20$) (see, Table~S2 of the \textit{supplement}). This strong bonding-antibonding is responsible for the semiconducting nature of this particular structure.
{The higher ICOHP suggests stronger covalent bonding, which induces the band-gap by stabilizing the valence states and destabilizing the conduction states by modifying hybridization. This is also consistent with the DFT calculated semiconducting behavior of \MG.}
This underscores the crucial interplay between the structural configuration and the electronic dynamics in these doped systems, highlighting how doping influences not only the local bonding environment but also the overall magnetic properties. 

{\par}Importantly, none of the Al atoms contribute to the magnetic properties of \MAl; rather, the Mn atoms are solely responsible for the ferrimagnetic behavior. This is further confirmed by Bader charge analysis (Table~\ref{tab:bader}), which reveals that Mn atoms in the system accumulate electrons, exhibiting a negative valency.
Let us first note that the ground-state electronic configurations of Mn and Al are [Ar]4$s^2$3$d^5$ and [Ne]3$s^2$3$p^1$, respectively. Each of the four Mn atoms in \MAl coordinates in a bicapped square antiprism configuration, surrounded by ten Al atoms. The Mn-Al bond lengths are all different. The Bader analysis reveals that Al atoms donate their 3$p$ electrons, exhibiting a positive valency. Mn atoms accumulate electrons, with 3.49 electrons at Mn$_1$ and $2.33$ electrons at Mn$_2$. The Mn-3$d^5$ electrons form a shell closure for the up-spin channel, which is responsible for the band-gap. Even after electron accumulation, both Mn$_1$ and Mn$_2$ still have unfilled 3$d$ states in the down-spin channel. As a result, the system becomes half-metallic.
The valency of Mn$_1$ is higher than Mn$_2$, a finding supported by the charge density plots in Figure~\ref{fig:MnAl}(c) and Figure~\ref{fig:MnAlGe1to6-band}(b). This difference in valency is a key factor in the ferrimagnetic nature of \MAl. In contrast,  The modifications of Mn-Al interactions due to Ge substitution in \MAl~leads a negative Ge valency, further altering the local electronic environment. This substitution not only introduces additional electrons but also modifies the local bonding and valency, particularly affecting the Mn atoms in the system. Interestingly, the valency of the non-substituted Al atoms remains unchanged compared to \MAl, with the only notable alteration occurring in the Mn atoms. Ultimately, in the \MG~ structure, the valencies of Mn$_1$ and Mn$_2$ become equal, which leads to the complete loss of the magnetic character, highlighting the significant impact of substitutional doping on the magnetic properties of the alloy.

\section{Conclusion}
The \MAl~ compound is intrinsically unique, with very few compounds exhibiting $10$-fold coordination in a bicapped square antiprismatic environment. The material-specific insights into \MAl, particularly concerning the collapse of the half-metallic gap under strain and the MIT upon Ge substitution in this low-symmetry intermetallic compound, offer valuable physical insights. Our study unequivocally demonstrates the robustness of the electronic and magnetic characteristics of the ferrimagnetic half-metallic \MAl~ compounds. One of the key findings is the significant role played by the Jahn-Teller distortion, which has a direct influence on the electronic properties of these structures, particularly given the notably lower crystal symmetry. The lack of symmetry is also prominently reflected in the magneto-crystalline anisotropy, reinforcing the idea that the crystal's structural characteristics are deeply intertwined with its magnetic behavior. The magnetization is observed to evolve steadily with applied strain, highlighting its inherent stability; however, the half-metallic band-gap is found to vanish beyond a $-2\%$ uniaxial strain. This observation suggests that while the half-metallic character is sensitive to strain, the magnetic properties are far more resilient, maintaining their integrity even as the electronic structure transitions. Thus, we can conclusively assert that the magnetic character of \MAl~ is more robust than its half-metallic nature under external strain, underlining the resilience of its ferrimagnetic properties. DFT and COHP analysis revealed how Ge substitution and uniaxial strain impact bonding interactions and magnetic behavior. While the half-metallic character of \MAl~ is sensitive to strain, the ferrimagnetic properties remain resilient. Additionally, Ge doping induces a MIT and alters the magnetic properties, highlighting the tunability of these materials for potential applications in spintronics and semiconductor technologies.
{The observed MIT in \MG~ is a distinctive consequence of site-specific substitution, which is closely tied to crystal symmetry. Although six Mn$_{4}$Al$_{11-x}$Ge$_x$ configurations were studied by substituting Al with Ge at six symmetry-inequivalent sites in the \MAl~ (P1) structure, we found that substitution at the ($000$) symmetry position is particularly critical for inducing the MIT. Therefore, we conclude that the band-gap in the Mn-Al-Ge system is not merely composition-dependent, but is instead strongly governed by the symmetry of the substitution site.}  
These findings offer valuable insights into designing advanced materials with tailored electronic and magnetic properties.

\section*{{Supporting Information}} It contains $12$ figures and $2$ tables used for supporting the discussions in main text.

\begin{acknowledgement}
PS was supported by the U.S. Department of Energy, Office of Science, Basic Energy Sciences, Materials Science and Engineering Division. The research was per- formed at the Ames National Laboratory, which is operated for the U.S. DOE by Iowa State University under Contract No. DE-AC02-07CH11358.
\end{acknowledgement}

\bibliographystyle{achemso}
\bibliography{Mn4Al11}

\providecommand{\latin}[1]{#1}
\makeatletter
\providecommand{\doi}
  {\begingroup\let\do\@makeother\dospecials
  \catcode`\{=1 \catcode`\}=2 \doi@aux}
\providecommand{\doi@aux}[1]{\endgroup\texttt{#1}}
\makeatother
\providecommand*\mcitethebibliography{\thebibliography}
\csname @ifundefined\endcsname{endmcitethebibliography}
  {\let\endmcitethebibliography\endthebibliography}{}
\begin{mcitethebibliography}{73}
\providecommand*\natexlab[1]{#1}
\providecommand*\mciteSetBstSublistMode[1]{}
\providecommand*\mciteSetBstMaxWidthForm[2]{}
\providecommand*\mciteBstWouldAddEndPuncttrue
  {\def\EndOfBibitem{\unskip.}}
\providecommand*\mciteBstWouldAddEndPunctfalse
  {\let\EndOfBibitem\relax}
\providecommand*\mciteSetBstMidEndSepPunct[3]{}
\providecommand*\mciteSetBstSublistLabelBeginEnd[3]{}
\providecommand*\EndOfBibitem{}
\mciteSetBstSublistMode{f}
\mciteSetBstMaxWidthForm{subitem}{(\alph{mcitesubitemcount})}
\mciteSetBstSublistLabelBeginEnd
  {\mcitemaxwidthsubitemform\space}
  {\relax}
  {\relax}

\bibitem[Stoloff \latin{et~al.}(2000)Stoloff, Liu, and
  Deevi]{stoloff2000emerging}
Stoloff,~N.; Liu,~C.; Deevi,~S. Emerging applications of intermetallics.
  \emph{Intermetallics} \textbf{2000}, \emph{8}, 1313--1320\relax
\mciteBstWouldAddEndPuncttrue
\mciteSetBstMidEndSepPunct{\mcitedefaultmidpunct}
{\mcitedefaultendpunct}{\mcitedefaultseppunct}\relax
\EndOfBibitem
\bibitem[Zhang \latin{et~al.}(2024)Zhang, Song, Luo, Shen, Hu, and
  Wang]{zhang2024recent}
Zhang,~Q.; Song,~M.; Luo,~G.; Shen,~T.; Hu,~H.; Wang,~D. Recent Advances of
  High-Entropy Intermetallics for Electrocatalysis. \emph{Chemistry of
  Materials} \textbf{2024}, \emph{36}, 10967--10985\relax
\mciteBstWouldAddEndPuncttrue
\mciteSetBstMidEndSepPunct{\mcitedefaultmidpunct}
{\mcitedefaultendpunct}{\mcitedefaultseppunct}\relax
\EndOfBibitem
\bibitem[Terada \latin{et~al.}(2002)Terada, Ohkubo, Mohri, and
  Suzuki]{terada2002thermal}
Terada,~Y.; Ohkubo,~K.; Mohri,~T.; Suzuki,~T. Thermal conductivity of
  intermetallic compounds with metallic bonding. \emph{Materials transactions}
  \textbf{2002}, \emph{43}, 3167--3176\relax
\mciteBstWouldAddEndPuncttrue
\mciteSetBstMidEndSepPunct{\mcitedefaultmidpunct}
{\mcitedefaultendpunct}{\mcitedefaultseppunct}\relax
\EndOfBibitem
\bibitem[Berry \latin{et~al.}(2017)Berry, Fu, Auffermann, Fecher, Schnelle,
  Serrano-Sanchez, Yue, Liang, and Felser]{berry2017enhancing}
Berry,~T.; Fu,~C.; Auffermann,~G.; Fecher,~G.~H.; Schnelle,~W.;
  Serrano-Sanchez,~F.; Yue,~Y.; Liang,~H.; Felser,~C. Enhancing thermoelectric
  performance of TiNiSn half-Heusler compounds via modulation doping.
  \emph{Chemistry of Materials} \textbf{2017}, \emph{29}, 7042--7048\relax
\mciteBstWouldAddEndPuncttrue
\mciteSetBstMidEndSepPunct{\mcitedefaultmidpunct}
{\mcitedefaultendpunct}{\mcitedefaultseppunct}\relax
\EndOfBibitem
\bibitem[Yuan \latin{et~al.}(2023)Yuan, Liu, Tang, Li, Huang, Zou, Yu, and
  Shui]{yuan2023honeycomb}
Yuan,~Y.; Liu,~X.; Tang,~W.; Li,~Z.; Huang,~G.; Zou,~H.; Yu,~R.; Shui,~J.
  Honeycomb ZrCo intermetallic for high performance hydrogen and hydrogen
  isotope storage. \emph{ACS Applied Materials \& Interfaces} \textbf{2023},
  \emph{15}, 3904--3911\relax
\mciteBstWouldAddEndPuncttrue
\mciteSetBstMidEndSepPunct{\mcitedefaultmidpunct}
{\mcitedefaultendpunct}{\mcitedefaultseppunct}\relax
\EndOfBibitem
\bibitem[Vaney \latin{et~al.}(2022)Vaney, Vignolle, Demourgues, Gaudin, Durand,
  Labrug{\`e}re, Bernardini, Cano, and Tenc{\'e}]{vaney2022topotactic}
Vaney,~J.-B.; Vignolle,~B.; Demourgues,~A.; Gaudin,~E.; Durand,~E.;
  Labrug{\`e}re,~C.; Bernardini,~F.; Cano,~A.; Tenc{\'e},~S. Topotactic
  fluorination of intermetallics as an efficient route towards quantum
  materials. \emph{Nature Communications} \textbf{2022}, \emph{13}, 1462\relax
\mciteBstWouldAddEndPuncttrue
\mciteSetBstMidEndSepPunct{\mcitedefaultmidpunct}
{\mcitedefaultendpunct}{\mcitedefaultseppunct}\relax
\EndOfBibitem
\bibitem[Adamski \latin{et~al.}(2024)Adamski, Zhang, Kaur, Chen, Liang, and
  Armbruster]{adamski2024selective}
Adamski,~P.; Zhang,~H.; Kaur,~S.; Chen,~X.; Liang,~C.; Armbruster,~M. Selective
  Hydrogenation of $\alpha$, $\beta$-Unsaturated Aldehydes Over Intermetallic
  Compounds- A Critical Review. \emph{Chemistry of Materials} \textbf{2024},
  \emph{36}, 10383--10407\relax
\mciteBstWouldAddEndPuncttrue
\mciteSetBstMidEndSepPunct{\mcitedefaultmidpunct}
{\mcitedefaultendpunct}{\mcitedefaultseppunct}\relax
\EndOfBibitem
\bibitem[Polmear \latin{et~al.}(2017)Polmear, StJohn, Nie, and
  Qian]{polmear2017light}
Polmear,~I.; StJohn,~D.; Nie,~J.-F.; Qian,~M. \emph{Light alloys: metallurgy of
  the light metals}; Butterworth-Heinemann, 2017\relax
\mciteBstWouldAddEndPuncttrue
\mciteSetBstMidEndSepPunct{\mcitedefaultmidpunct}
{\mcitedefaultendpunct}{\mcitedefaultseppunct}\relax
\EndOfBibitem
\bibitem[Wei \latin{et~al.}(2009)Wei, Wei, Du, and Hu]{wei2009}
Wei,~K.~X.; Wei,~W.; Du,~Q.~B.; Hu,~J. Microstructure and tensile properties of
  Al--Mn alloy processed by accumulative roll bonding. \emph{Materials Science
  and Engineering: A} \textbf{2009}, \emph{525}, 55--59\relax
\mciteBstWouldAddEndPuncttrue
\mciteSetBstMidEndSepPunct{\mcitedefaultmidpunct}
{\mcitedefaultendpunct}{\mcitedefaultseppunct}\relax
\EndOfBibitem
\bibitem[Nam and Lee(2000)Nam, and Lee]{nam2000effect}
Nam,~S.~W.; Lee,~D.~H. The effect of Mn on the mechanical behavior of Al
  alloys. \emph{Metals and materials} \textbf{2000}, \emph{6}, 13--16\relax
\mciteBstWouldAddEndPuncttrue
\mciteSetBstMidEndSepPunct{\mcitedefaultmidpunct}
{\mcitedefaultendpunct}{\mcitedefaultseppunct}\relax
\EndOfBibitem
\bibitem[Shen \latin{et~al.}(2023)Shen, Hu, Liu, and Hu]{shen2023mn}
Shen,~W.; Hu,~A.; Liu,~S.; Hu,~H. Al-Mn alloys for electrical applications: a
  review. \emph{Journal of Alloys and Metallurgical Systems} \textbf{2023},
  100008\relax
\mciteBstWouldAddEndPuncttrue
\mciteSetBstMidEndSepPunct{\mcitedefaultmidpunct}
{\mcitedefaultendpunct}{\mcitedefaultseppunct}\relax
\EndOfBibitem
\bibitem[McAlister and Murray(1987)McAlister, and Murray]{mcalister1987mn}
McAlister,~A.; Murray,~J. The (Al- Mn) aluminum-manganese system. \emph{Journal
  of Phase Equilibria} \textbf{1987}, \emph{8}, 438--447\relax
\mciteBstWouldAddEndPuncttrue
\mciteSetBstMidEndSepPunct{\mcitedefaultmidpunct}
{\mcitedefaultendpunct}{\mcitedefaultseppunct}\relax
\EndOfBibitem
\bibitem[Liu \latin{et~al.}(1999)Liu, Ohnuma, Kainuma, and
  Ishida]{liu1999thermodynamic}
Liu,~X.; Ohnuma,~I.; Kainuma,~R.; Ishida,~K. Thermodynamic assessment of the
  Aluminum-Manganese (Al-Mn) binary phase diagram. \emph{Journal of phase
  equilibria} \textbf{1999}, \emph{20}, 45--56\relax
\mciteBstWouldAddEndPuncttrue
\mciteSetBstMidEndSepPunct{\mcitedefaultmidpunct}
{\mcitedefaultendpunct}{\mcitedefaultseppunct}\relax
\EndOfBibitem
\bibitem[Gorshenkov \latin{et~al.}(2020)Gorshenkov, Karpenkov, Sundeev,
  Cheverikin, and Shchetinin]{gorshenkov2020magnetic}
Gorshenkov,~M.; Karpenkov,~D.; Sundeev,~R.; Cheverikin,~V.; Shchetinin,~I.
  Magnetic properties of Mn-Al alloy after HPT deformation. \emph{Materials
  Letters} \textbf{2020}, \emph{272}, 127864\relax
\mciteBstWouldAddEndPuncttrue
\mciteSetBstMidEndSepPunct{\mcitedefaultmidpunct}
{\mcitedefaultendpunct}{\mcitedefaultseppunct}\relax
\EndOfBibitem
\bibitem[Du \latin{et~al.}(2007)Du, Wang, Zhao, Schuster, Weitzer,
  Schmid-Fetzer, Ohno, Xu, Liu, Shang, \latin{et~al.}
  others]{du2007reassessment}
Du,~Y.; Wang,~J.; Zhao,~J.; Schuster,~J.~C.; Weitzer,~F.; Schmid-Fetzer,~R.;
  Ohno,~M.; Xu,~H.; Liu,~Z.-k.; Shang,~S., \latin{et~al.}  Reassessment of the
  Al--Mn system and a thermodynamic description of the Al--Mg--Mn system.
  \emph{International journal of materials research} \textbf{2007}, \emph{98},
  855--871\relax
\mciteBstWouldAddEndPuncttrue
\mciteSetBstMidEndSepPunct{\mcitedefaultmidpunct}
{\mcitedefaultendpunct}{\mcitedefaultseppunct}\relax
\EndOfBibitem
\bibitem[Murray \latin{et~al.}(1987)Murray, McAlister, Schaefer, Bendersky,
  Biancaniello, and Moffat]{murray1987stable}
Murray,~J.; McAlister,~A.; Schaefer,~R.; Bendersky,~L.; Biancaniello,~F.;
  Moffat,~D. Stable and metastable phase equilibria in the Al-Mn system.
  \emph{Metallurgical Transactions A} \textbf{1987}, \emph{18}, 385--392\relax
\mciteBstWouldAddEndPuncttrue
\mciteSetBstMidEndSepPunct{\mcitedefaultmidpunct}
{\mcitedefaultendpunct}{\mcitedefaultseppunct}\relax
\EndOfBibitem
\bibitem[Shukla and Pelton(2009)Shukla, and Pelton]{shukla2009thermodynamic}
Shukla,~A.; Pelton,~A.~D. Thermodynamic assessment of the Al-Mn and Mg-Al-Mn
  systems. \emph{Journal of phase equilibria and diffusion} \textbf{2009},
  \emph{30}, 28--39\relax
\mciteBstWouldAddEndPuncttrue
\mciteSetBstMidEndSepPunct{\mcitedefaultmidpunct}
{\mcitedefaultendpunct}{\mcitedefaultseppunct}\relax
\EndOfBibitem
\bibitem[Kumar \latin{et~al.}(1986)Kumar, Sahoo, and Athithan]{kumar1986}
Kumar,~V.; Sahoo,~D.; Athithan,~G. Characterization and decoration of the
  two-dimensional Penrose lattice. \emph{Physical Review B} \textbf{1986},
  \emph{34}, 6924\relax
\mciteBstWouldAddEndPuncttrue
\mciteSetBstMidEndSepPunct{\mcitedefaultmidpunct}
{\mcitedefaultendpunct}{\mcitedefaultseppunct}\relax
\EndOfBibitem
\bibitem[Mihalkovi{\v{c}} \latin{et~al.}(1996)Mihalkovi{\v{c}}, Zhu, Henley,
  and Phillips]{mihalkovivc1996}
Mihalkovi{\v{c}},~M.; Zhu,~W.-J.; Henley,~C.; Phillips,~R. Icosahedral
  quasicrystal decoration models. II. Optimization under realistic Al-Mn
  potentials. \emph{Physical Review B} \textbf{1996}, \emph{53}, 9021\relax
\mciteBstWouldAddEndPuncttrue
\mciteSetBstMidEndSepPunct{\mcitedefaultmidpunct}
{\mcitedefaultendpunct}{\mcitedefaultseppunct}\relax
\EndOfBibitem
\bibitem[Kontio \latin{et~al.}(1980)Kontio, Stevens, Coppens, Brown, Dwight,
  and Williams]{kontio1980}
Kontio,~A.; Stevens,~E.; Coppens,~P.; Brown,~R.; Dwight,~A.; Williams,~J. New
  investigation of the structure of Mn4Al11. \emph{Acta Crystallographica
  Section B: Structural Crystallography and Crystal Chemistry} \textbf{1980},
  \emph{36}, 435--436\relax
\mciteBstWouldAddEndPuncttrue
\mciteSetBstMidEndSepPunct{\mcitedefaultmidpunct}
{\mcitedefaultendpunct}{\mcitedefaultseppunct}\relax
\EndOfBibitem
\bibitem[Bland(1958)]{bland1958}
Bland,~J. Studies of aluminum-rich alloys with the transition metals manganese
  and tungsten. II. The crystal structure of $\delta$ (Mn-Al)-Mn4A111.
  \emph{Acta Crystallographica} \textbf{1958}, \emph{11}, 236--244\relax
\mciteBstWouldAddEndPuncttrue
\mciteSetBstMidEndSepPunct{\mcitedefaultmidpunct}
{\mcitedefaultendpunct}{\mcitedefaultseppunct}\relax
\EndOfBibitem
\bibitem[Schaefer \latin{et~al.}(1986)Schaefer, Biancaniello, and
  Cahn]{schaefer1986formation}
Schaefer,~R.; Biancaniello,~F.; Cahn,~J. Formation and stability range of the G
  phase in the Al- Mn system. \emph{Scripta metallurgica} \textbf{1986},
  \emph{20}, 1439--1444\relax
\mciteBstWouldAddEndPuncttrue
\mciteSetBstMidEndSepPunct{\mcitedefaultmidpunct}
{\mcitedefaultendpunct}{\mcitedefaultseppunct}\relax
\EndOfBibitem
\bibitem[Godecke and Koster(1971)Godecke, and Koster]{godecke1971supplement}
Godecke,~T.; Koster,~W. A supplement to the constitution of the
  Aluminum-Manganese System. \emph{Z. Metallkunde} \textbf{1971}, \emph{62},
  727--732\relax
\mciteBstWouldAddEndPuncttrue
\mciteSetBstMidEndSepPunct{\mcitedefaultmidpunct}
{\mcitedefaultendpunct}{\mcitedefaultseppunct}\relax
\EndOfBibitem
\bibitem[Grushko and Balanetskyy(2008)Grushko, and
  Balanetskyy]{grushko2008study}
Grushko,~B.; Balanetskyy,~S. A study of phase equilibria in the Al-rich part of
  the Al-Mn alloy system. \emph{International journal of materials research}
  \textbf{2008}, \emph{99}, 1319--1323\relax
\mciteBstWouldAddEndPuncttrue
\mciteSetBstMidEndSepPunct{\mcitedefaultmidpunct}
{\mcitedefaultendpunct}{\mcitedefaultseppunct}\relax
\EndOfBibitem
\bibitem[Caplin and Rizzuto(1968)Caplin, and Rizzuto]{caplin1968}
Caplin,~A.; Rizzuto,~C. Anomalies in the electrical resistance of Al: Mn and
  Al: Cr alloys. \emph{Physical Review Letters} \textbf{1968}, \emph{21},
  746\relax
\mciteBstWouldAddEndPuncttrue
\mciteSetBstMidEndSepPunct{\mcitedefaultmidpunct}
{\mcitedefaultendpunct}{\mcitedefaultseppunct}\relax
\EndOfBibitem
\bibitem[Bak(1985)]{bak1985}
Bak,~P. Phenomenological theory of icosahedral incommensurate ("
  quasiperiodic") order in Mn-Al alloys. \emph{Physical review letters}
  \textbf{1985}, \emph{54}, 1517\relax
\mciteBstWouldAddEndPuncttrue
\mciteSetBstMidEndSepPunct{\mcitedefaultmidpunct}
{\mcitedefaultendpunct}{\mcitedefaultseppunct}\relax
\EndOfBibitem
\bibitem[Zhao \latin{et~al.}(2015)Zhao, Hozumi, LeClair, Mankey, and
  Suzuki]{zhao2015}
Zhao,~S.; Hozumi,~T.; LeClair,~P.; Mankey,~G.; Suzuki,~T. Magnetic Anisotropy
  of $\tau$-MnAl Thin Films. \emph{IEEE Transactions on Magnetics}
  \textbf{2015}, \emph{51}, 1--4\relax
\mciteBstWouldAddEndPuncttrue
\mciteSetBstMidEndSepPunct{\mcitedefaultmidpunct}
{\mcitedefaultendpunct}{\mcitedefaultseppunct}\relax
\EndOfBibitem
\bibitem[Aghili \latin{et~al.}(2016)Aghili, Beiranvand, Elahi, and
  Abolhasani]{aghili2016half}
Aghili,~S.; Beiranvand,~R.; Elahi,~S.; Abolhasani,~M. Half-metallic
  ferromagnetism in Mn-doped zigzag AlN nanoribbon from first-principles.
  \emph{Journal of Magnetism and Magnetic Materials} \textbf{2016}, \emph{420},
  122--128\relax
\mciteBstWouldAddEndPuncttrue
\mciteSetBstMidEndSepPunct{\mcitedefaultmidpunct}
{\mcitedefaultendpunct}{\mcitedefaultseppunct}\relax
\EndOfBibitem
\bibitem[Yao \latin{et~al.}(2016)Yao, Li, and Zhou]{yao2016first}
Yao,~L.; Li,~K.; Zhou,~N. First-principles study of Mn adsorption on Al4C3 (0 0
  0 1) surface. \emph{Applied Surface Science} \textbf{2016}, \emph{363},
  168--172\relax
\mciteBstWouldAddEndPuncttrue
\mciteSetBstMidEndSepPunct{\mcitedefaultmidpunct}
{\mcitedefaultendpunct}{\mcitedefaultseppunct}\relax
\EndOfBibitem
\bibitem[Bohlmann \latin{et~al.}(1981)Bohlmann, Koo, and Wise]{bohlmann1981}
Bohlmann,~M.; Koo,~J.; Wise,~J. Mn-Al-C for permanent magnets. \emph{Journal of
  Applied Physics} \textbf{1981}, \emph{52}, 2542--2543\relax
\mciteBstWouldAddEndPuncttrue
\mciteSetBstMidEndSepPunct{\mcitedefaultmidpunct}
{\mcitedefaultendpunct}{\mcitedefaultseppunct}\relax
\EndOfBibitem
\bibitem[K{\=o}no(1958)]{kono1958}
K{\=o}no,~H. On the ferromagnetic phase in manganese-aluminum system.
  \emph{Journal of the Physical Society of Japan} \textbf{1958}, \emph{13},
  1444--1451\relax
\mciteBstWouldAddEndPuncttrue
\mciteSetBstMidEndSepPunct{\mcitedefaultmidpunct}
{\mcitedefaultendpunct}{\mcitedefaultseppunct}\relax
\EndOfBibitem
\bibitem[Van~Roy \latin{et~al.}(1995)Van~Roy, Bender, Bruynseraede, De~Boeck,
  and Borghs]{van1995}
Van~Roy,~W.; Bender,~H.; Bruynseraede,~C.; De~Boeck,~J.; Borghs,~G. Degree of
  order and magnetic properties of $\tau$-MnAl films. \emph{Journal of
  magnetism and magnetic materials} \textbf{1995}, \emph{148}, 97--98\relax
\mciteBstWouldAddEndPuncttrue
\mciteSetBstMidEndSepPunct{\mcitedefaultmidpunct}
{\mcitedefaultendpunct}{\mcitedefaultseppunct}\relax
\EndOfBibitem
\bibitem[Zhu \latin{et~al.}(2016)Zhu, Nie, Xiong, Schlottmann, and
  Zhao]{zhu2016}
Zhu,~L.; Nie,~S.; Xiong,~P.; Schlottmann,~P.; Zhao,~J. Orbital two-channel
  Kondo effect in epitaxial ferromagnetic L 10-MnAl films. \emph{Nature
  communications} \textbf{2016}, \emph{7}, 10817\relax
\mciteBstWouldAddEndPuncttrue
\mciteSetBstMidEndSepPunct{\mcitedefaultmidpunct}
{\mcitedefaultendpunct}{\mcitedefaultseppunct}\relax
\EndOfBibitem
\bibitem[Sakuma(1994)]{sakuma1994}
Sakuma,~A. Electronic structure and magnetocrystalline anisotropy energy of
  MnAl. \emph{Journal of the Physical Society of Japan} \textbf{1994},
  \emph{63}, 1422--1428\relax
\mciteBstWouldAddEndPuncttrue
\mciteSetBstMidEndSepPunct{\mcitedefaultmidpunct}
{\mcitedefaultendpunct}{\mcitedefaultseppunct}\relax
\EndOfBibitem
\bibitem[M{\^o}ri and Mitsui(1968)M{\^o}ri, and Mitsui]{mori1968localized}
M{\^o}ri,~N.; Mitsui,~T. Localized magnetic moments and Pauling valence in
  manganese metal, some 3d-transition alloys and intermetallic compounds.
  \emph{Journal of the Physical Society of Japan} \textbf{1968}, \emph{25},
  82--88\relax
\mciteBstWouldAddEndPuncttrue
\mciteSetBstMidEndSepPunct{\mcitedefaultmidpunct}
{\mcitedefaultendpunct}{\mcitedefaultseppunct}\relax
\EndOfBibitem
\bibitem[Price(1978)]{price1978indirect}
Price,~D. Indirect magnetic coupling between local moments in metals.
  \emph{Journal of Physics F: Metal Physics} \textbf{1978}, \emph{8}, 933\relax
\mciteBstWouldAddEndPuncttrue
\mciteSetBstMidEndSepPunct{\mcitedefaultmidpunct}
{\mcitedefaultendpunct}{\mcitedefaultseppunct}\relax
\EndOfBibitem
\bibitem[Park \latin{et~al.}(2010)Park, Hong, Bae, Lee, Jalli, Abo, Neveu, Kim,
  Choi, and Lee]{park2010}
Park,~J.; Hong,~Y.; Bae,~S.; Lee,~J.; Jalli,~J.; Abo,~G.; Neveu,~N.; Kim,~S.;
  Choi,~C.; Lee,~J. Saturation magnetization and crystalline anisotropy
  calculations for MnAl permanent magnet. \emph{Journal of Applied Physics}
  \textbf{2010}, \emph{107}, 09A731\relax
\mciteBstWouldAddEndPuncttrue
\mciteSetBstMidEndSepPunct{\mcitedefaultmidpunct}
{\mcitedefaultendpunct}{\mcitedefaultseppunct}\relax
\EndOfBibitem
\bibitem[Kuo \latin{et~al.}(1992)Kuo, Yao, Huang, and Chen]{kuo1992}
Kuo,~P.; Yao,~Y.; Huang,~J.; Chen,~C. Fabrication and magnetic properties of
  manganese-aluminium permanent magnets. \emph{Journal of magnetism and
  magnetic materials} \textbf{1992}, \emph{115}, 183--186\relax
\mciteBstWouldAddEndPuncttrue
\mciteSetBstMidEndSepPunct{\mcitedefaultmidpunct}
{\mcitedefaultendpunct}{\mcitedefaultseppunct}\relax
\EndOfBibitem
\bibitem[Tsai \latin{et~al.}(1988)Tsai, Inoue, Masumoto, and Kataoka]{tsai1988}
Tsai,~A.-P.; Inoue,~A.; Masumoto,~T.; Kataoka,~N. Al-Ge-Mn and Al-Cu-Ge-Mn
  quasi-crystals with coercivity at room temperature. \emph{Japanese journal of
  applied physics} \textbf{1988}, \emph{27}, L2252\relax
\mciteBstWouldAddEndPuncttrue
\mciteSetBstMidEndSepPunct{\mcitedefaultmidpunct}
{\mcitedefaultendpunct}{\mcitedefaultseppunct}\relax
\EndOfBibitem
\bibitem[Ido \latin{et~al.}(1984)Ido, Kamimura, and Shirakawa]{ido1984}
Ido,~H.; Kamimura,~T.; Shirakawa,~K. Magnetic properties of Mn1- xMxAlGe (M= 3d
  metals). \emph{Journal of applied physics} \textbf{1984}, \emph{55},
  2365--2366\relax
\mciteBstWouldAddEndPuncttrue
\mciteSetBstMidEndSepPunct{\mcitedefaultmidpunct}
{\mcitedefaultendpunct}{\mcitedefaultseppunct}\relax
\EndOfBibitem
\bibitem[Lalla \latin{et~al.}(1992)Lalla, Tiwari, and Srivastava]{lalla1992}
Lalla,~N.; Tiwari,~R.; Srivastava,~O. Coherent orientation relationship among
  various quasicrystalline and crystalline phases in rapidly solidified
  Al78Mn20Ge2 alloy. \emph{Journal of materials research} \textbf{1992},
  \emph{7}, 53--61\relax
\mciteBstWouldAddEndPuncttrue
\mciteSetBstMidEndSepPunct{\mcitedefaultmidpunct}
{\mcitedefaultendpunct}{\mcitedefaultseppunct}\relax
\EndOfBibitem
\bibitem[Schaefer and Bendersky(1986)Schaefer, and Bendersky]{schaefer1986}
Schaefer,~R.; Bendersky,~L. Replacement of icosahedral Al-Mn by decagonal
  phase. \emph{Scripta metallurgica} \textbf{1986}, \emph{20}, 745--750\relax
\mciteBstWouldAddEndPuncttrue
\mciteSetBstMidEndSepPunct{\mcitedefaultmidpunct}
{\mcitedefaultendpunct}{\mcitedefaultseppunct}\relax
\EndOfBibitem
\bibitem[Sarvar \latin{et~al.}(2024)Sarvar, Nodir, Mardonov, Saydumarov,
  Kulmuradov, and Boltaeva]{sarvar2024}
Sarvar,~T.; Nodir,~T.; Mardonov,~U.; Saydumarov,~B.; Kulmuradov,~D.;
  Boltaeva,~M. EFFECTS OF GERMANIUM (GE) ON HARDNESS AND MICROSTRUCTURE OF
  AL-MG, AL-CU, AL-MN SYSTEM ALLOYS. \emph{International Journal of
  Mechatronics and Applied Mechanics} \textbf{2024}, 179--184\relax
\mciteBstWouldAddEndPuncttrue
\mciteSetBstMidEndSepPunct{\mcitedefaultmidpunct}
{\mcitedefaultendpunct}{\mcitedefaultseppunct}\relax
\EndOfBibitem
\bibitem[Perdew(2008)]{PBEsol}
Perdew,~J. JP Perdew, A. Ruzsinszky, GI Csonka, OA Vydrov, GE Scuseria, LA
  Constantin, X. Zhou, and K. Burke, Phys. Rev. Lett. 100, 136406 (2008).
  \emph{Phys. Rev. Lett.} \textbf{2008}, \emph{100}, 136406\relax
\mciteBstWouldAddEndPuncttrue
\mciteSetBstMidEndSepPunct{\mcitedefaultmidpunct}
{\mcitedefaultendpunct}{\mcitedefaultseppunct}\relax
\EndOfBibitem
\bibitem[Giannozzi \latin{et~al.}(2017)Giannozzi, Andreussi, Brumme, Bunau,
  Nardelli, Calandra, Car, Cavazzoni, Ceresoli, Cococcioni, Colonna, Carnimeo,
  Corso, de~Gironcoli, Delugas, DiStasio, Ferretti, Floris, Fratesi, Fugallo,
  Gebauer, Gerstmann, Giustino, Gorni, Jia, Kawamura, Ko, Kokalj,
  K\"u{\c{c}}\"ukbenli, Lazzeri, Marsili, Marzari, Mauri, Nguyen, Nguyen, de-la
  Roza, Paulatto, Ponc{\'{e}}, Rocca, Sabatini, Santra, Schlipf, Seitsonen,
  Smogunov, Timrov, Thonhauser, Umari, Vast, Wu, and Baroni]{QE}
Giannozzi,~P. \latin{et~al.}  Advanced capabilities for materials modelling
  with Quantum {ESPRESSO}. \emph{Journal of Physics: Condensed Matter}
  \textbf{2017}, \emph{29}, 465901\relax
\mciteBstWouldAddEndPuncttrue
\mciteSetBstMidEndSepPunct{\mcitedefaultmidpunct}
{\mcitedefaultendpunct}{\mcitedefaultseppunct}\relax
\EndOfBibitem
\bibitem[Marzari \latin{et~al.}(1999)Marzari, Vanderbilt, De~Vita, and
  Payne]{marzari1999}
Marzari,~N.; Vanderbilt,~D.; De~Vita,~A.; Payne,~M. Thermal contraction and
  disordering of the Al (110) surface. \emph{Physical review letters}
  \textbf{1999}, \emph{82}, 3296\relax
\mciteBstWouldAddEndPuncttrue
\mciteSetBstMidEndSepPunct{\mcitedefaultmidpunct}
{\mcitedefaultendpunct}{\mcitedefaultseppunct}\relax
\EndOfBibitem
\bibitem[Togo(2023)]{togo2023first}
Togo,~A. First-principles phonon calculations with phonopy and phono3py.
  \emph{Journal of the Physical Society of Japan} \textbf{2023}, \emph{92},
  012001\relax
\mciteBstWouldAddEndPuncttrue
\mciteSetBstMidEndSepPunct{\mcitedefaultmidpunct}
{\mcitedefaultendpunct}{\mcitedefaultseppunct}\relax
\EndOfBibitem
\bibitem[Baroni \latin{et~al.}(2001)Baroni, De~Gironcoli, Dal~Corso, and
  Giannozzi]{baroni2001phonons}
Baroni,~S.; De~Gironcoli,~S.; Dal~Corso,~A.; Giannozzi,~P. Phonons and related
  crystal properties from density-functional perturbation theory. \emph{Reviews
  of modern Physics} \textbf{2001}, \emph{73}, 515\relax
\mciteBstWouldAddEndPuncttrue
\mciteSetBstMidEndSepPunct{\mcitedefaultmidpunct}
{\mcitedefaultendpunct}{\mcitedefaultseppunct}\relax
\EndOfBibitem
\bibitem[Yalameha \latin{et~al.}(2022)Yalameha, Nourbakhsh, and
  Vashaee]{elatools}
Yalameha,~S.; Nourbakhsh,~Z.; Vashaee,~D. ElATools: A tool for analyzing
  anisotropic elastic properties of the 2D and 3D materials. \emph{Computer
  Physics Communications} \textbf{2022}, \emph{271}, 108195\relax
\mciteBstWouldAddEndPuncttrue
\mciteSetBstMidEndSepPunct{\mcitedefaultmidpunct}
{\mcitedefaultendpunct}{\mcitedefaultseppunct}\relax
\EndOfBibitem
\bibitem[Momma and Izumi(2008)Momma, and Izumi]{vesta}
Momma,~K.; Izumi,~F. VESTA: a three-dimensional visualization system for
  electronic and structural analysis. \emph{Journal of Applied crystallography}
  \textbf{2008}, \emph{41}, 653--658\relax
\mciteBstWouldAddEndPuncttrue
\mciteSetBstMidEndSepPunct{\mcitedefaultmidpunct}
{\mcitedefaultendpunct}{\mcitedefaultseppunct}\relax
\EndOfBibitem
\bibitem[Nelson \latin{et~al.}(2020)Nelson, Ertural, George, Deringer, Hautier,
  and Dronskowski]{nelson2020}
Nelson,~R.; Ertural,~C.; George,~J.; Deringer,~V.~L.; Hautier,~G.;
  Dronskowski,~R. LOBSTER: Local orbital projections, atomic charges, and
  chemical-bonding analysis from projector-augmented-wave-based
  density-functional theory. \emph{Journal of Computational Chemistry}
  \textbf{2020}, \emph{41}, 1931--1940\relax
\mciteBstWouldAddEndPuncttrue
\mciteSetBstMidEndSepPunct{\mcitedefaultmidpunct}
{\mcitedefaultendpunct}{\mcitedefaultseppunct}\relax
\EndOfBibitem
\bibitem[Pearson \latin{et~al.}(1958)Pearson, \latin{et~al.}
  others]{pearson1958lattice}
Pearson,~W., \latin{et~al.}  Lattice spacings and structures of metals and
  alloys. \emph{Vols. I and II (Pergamon Press, Oxford, 1964, 1967)}
  \textbf{1958}, \relax
\mciteBstWouldAddEndPunctfalse
\mciteSetBstMidEndSepPunct{\mcitedefaultmidpunct}
{}{\mcitedefaultseppunct}\relax
\EndOfBibitem
\bibitem[Singh \latin{et~al.}(2024)Singh, Del~Rose, Mudryk, Pecharsky, and
  Johnson]{PhysRevB.109.064207}
Singh,~P.; Del~Rose,~T.; Mudryk,~Y.; Pecharsky,~V.~K.; Johnson,~D.~D. Designed
  metal-insulator transition in low-symmetry magnetic intermetallics.
  \emph{Phys. Rev. B} \textbf{2024}, \emph{109}, 064207\relax
\mciteBstWouldAddEndPuncttrue
\mciteSetBstMidEndSepPunct{\mcitedefaultmidpunct}
{\mcitedefaultendpunct}{\mcitedefaultseppunct}\relax
\EndOfBibitem
\bibitem[Belin-Ferr{\'e} \latin{et~al.}(2010)Belin-Ferr{\'e}, Dankhazi,
  Fontaine, de~Weerd, and Dubois]{belin2010}
Belin-Ferr{\'e},~E.; Dankhazi,~Z.; Fontaine,~M.-F.; de~Weerd,~M.-C.;
  Dubois,~J.~M. Electronic densities of states and pseudo-gaps in Al-based
  complex intermetallics. \emph{Croatica Chemica Acta} \textbf{2010},
  \emph{83}, 55--58\relax
\mciteBstWouldAddEndPuncttrue
\mciteSetBstMidEndSepPunct{\mcitedefaultmidpunct}
{\mcitedefaultendpunct}{\mcitedefaultseppunct}\relax
\EndOfBibitem
\bibitem[{\"O}zdog \latin{et~al.}(2007){\"O}zdog, Galanakis,
  {\c{S}}a{\c{s}}{\i}og, and Akta{\c{s}}]{ozdog2007}
{\"O}zdog,~K.; Galanakis,~I.; {\c{S}}a{\c{s}}{\i}og,~E.; Akta{\c{s}},~B.
  Defects-driven appearance of half-metallic ferrimagnetism in Co--Mn-based
  Heusler alloys. \emph{Solid state communications} \textbf{2007}, \emph{142},
  492--497\relax
\mciteBstWouldAddEndPuncttrue
\mciteSetBstMidEndSepPunct{\mcitedefaultmidpunct}
{\mcitedefaultendpunct}{\mcitedefaultseppunct}\relax
\EndOfBibitem
\bibitem[Ostler \latin{et~al.}(2011)Ostler, Evans, Chantrell, Atxitia,
  Chubykalo-Fesenko, Radu, Abrudan, Radu, Tsukamoto, Itoh, \latin{et~al.}
  others]{ostler2011}
Ostler,~T.~A.; Evans,~R.~F.; Chantrell,~R.~W.; Atxitia,~U.;
  Chubykalo-Fesenko,~O.; Radu,~I.; Abrudan,~R.; Radu,~F.; Tsukamoto,~A.;
  Itoh,~A., \latin{et~al.}  Crystallographically amorphous ferrimagnetic
  alloys: Comparing a localized atomistic spin model with experiments.
  \emph{Physical Review B} \textbf{2011}, \emph{84}, 024407\relax
\mciteBstWouldAddEndPuncttrue
\mciteSetBstMidEndSepPunct{\mcitedefaultmidpunct}
{\mcitedefaultendpunct}{\mcitedefaultseppunct}\relax
\EndOfBibitem
\bibitem[Je \latin{et~al.}(2018)Je, Rojas-Sanchez, Pham, Vallobra, Malinowski,
  Lacour, Fache, Cyrille, Kim, Choe, \latin{et~al.} others]{je2018}
Je,~S.-G.; Rojas-Sanchez,~J.-C.; Pham,~T.~H.; Vallobra,~P.; Malinowski,~G.;
  Lacour,~D.; Fache,~T.; Cyrille,~M.-C.; Kim,~D.-Y.; Choe,~S.-B.,
  \latin{et~al.}  Spin-orbit torque-induced switching in ferrimagnetic alloys:
  Experiments and modeling. \emph{Applied Physics Letters} \textbf{2018},
  \emph{112}, 062401\relax
\mciteBstWouldAddEndPuncttrue
\mciteSetBstMidEndSepPunct{\mcitedefaultmidpunct}
{\mcitedefaultendpunct}{\mcitedefaultseppunct}\relax
\EndOfBibitem
\bibitem[Jiang \latin{et~al.}(2001)Jiang, Venkatesan, and Coey]{jiang2001}
Jiang,~C.; Venkatesan,~M.; Coey,~J. Transport and magnetic properties of
  Mn2VAl: Search for half-metallicity. \emph{Solid state communications}
  \textbf{2001}, \emph{118}, 513--516\relax
\mciteBstWouldAddEndPuncttrue
\mciteSetBstMidEndSepPunct{\mcitedefaultmidpunct}
{\mcitedefaultendpunct}{\mcitedefaultseppunct}\relax
\EndOfBibitem
\bibitem[Jum’h \latin{et~al.}(2019)Jum’h, S{\^a}ad~Essaoud, Baaziz,
  Charifi, and Telfah]{jum2019}
Jum’h,~I.; S{\^a}ad~Essaoud,~S.; Baaziz,~H.; Charifi,~Z.; Telfah,~A.
  Electronic and magnetic structure and elastic and thermal properties of
  Mn2-based full Heusler alloys. \emph{Journal of Superconductivity and Novel
  Magnetism} \textbf{2019}, \emph{32}, 3915--3926\relax
\mciteBstWouldAddEndPuncttrue
\mciteSetBstMidEndSepPunct{\mcitedefaultmidpunct}
{\mcitedefaultendpunct}{\mcitedefaultseppunct}\relax
\EndOfBibitem
\bibitem[Dunlop and Gr{\"u}ner(1976)Dunlop, and Gr{\"u}ner]{dunlop1976}
Dunlop,~J.; Gr{\"u}ner,~G. One-dimensional effects in the intermetallic
  compound Al11Mn4. \emph{Solid State Communications} \textbf{1976}, \emph{18},
  827--829\relax
\mciteBstWouldAddEndPuncttrue
\mciteSetBstMidEndSepPunct{\mcitedefaultmidpunct}
{\mcitedefaultendpunct}{\mcitedefaultseppunct}\relax
\EndOfBibitem
\bibitem[Jamer \latin{et~al.}(2017)Jamer, Wang, Stephen, McDonald, Grutter,
  Sterbinsky, Arena, Borchers, Kirby, Lewis, \latin{et~al.} others]{jamer2017}
Jamer,~M.~E.; Wang,~Y.~J.; Stephen,~G.~M.; McDonald,~I.~J.; Grutter,~A.~J.;
  Sterbinsky,~G.~E.; Arena,~D.~A.; Borchers,~J.~A.; Kirby,~B.~J.; Lewis,~L.~H.,
  \latin{et~al.}  Compensated ferrimagnetism in the zero-moment Heusler alloy
  Mn 3 Al. \emph{Physical Review Applied} \textbf{2017}, \emph{7}, 064036\relax
\mciteBstWouldAddEndPuncttrue
\mciteSetBstMidEndSepPunct{\mcitedefaultmidpunct}
{\mcitedefaultendpunct}{\mcitedefaultseppunct}\relax
\EndOfBibitem
\bibitem[Luo \latin{et~al.}(2008)Luo, Zhu, Ma, Xu, Zhu, Jiang, Xu, and
  Wu]{luo2008}
Luo,~H.; Zhu,~Z.; Ma,~L.; Xu,~S.; Zhu,~X.; Jiang,~C.; Xu,~H.; Wu,~G. Effect of
  site preference of 3d atoms on the electronic structure and half-metallicity
  of Heusler alloy Mn2YAl. \emph{Journal of Physics D: Applied Physics}
  \textbf{2008}, \emph{41}, 055010\relax
\mciteBstWouldAddEndPuncttrue
\mciteSetBstMidEndSepPunct{\mcitedefaultmidpunct}
{\mcitedefaultendpunct}{\mcitedefaultseppunct}\relax
\EndOfBibitem
\bibitem[Ouardi \latin{et~al.}(2013)Ouardi, Fecher, Felser, and
  K{\"u}bler]{ouardi2013}
Ouardi,~S.; Fecher,~G.~H.; Felser,~C.; K{\"u}bler,~J. Realization of spin
  gapless semiconductors: The Heusler compound Mn 2 CoAl. \emph{Physical review
  letters} \textbf{2013}, \emph{110}, 100401\relax
\mciteBstWouldAddEndPuncttrue
\mciteSetBstMidEndSepPunct{\mcitedefaultmidpunct}
{\mcitedefaultendpunct}{\mcitedefaultseppunct}\relax
\EndOfBibitem
\bibitem[Kourov \latin{et~al.}(2015)Kourov, Marchenkov, Korolev, Stashkova,
  Emel’yanova, and Weber]{kourov2015}
Kourov,~N.; Marchenkov,~V.; Korolev,~A.; Stashkova,~L.; Emel’yanova,~S.;
  Weber,~H. Specific features of the properties of half-metallic ferromagnetic
  Heusler alloys Fe 2 MnAl, Fe 2 MnSi, and Co 2 MnAl. \emph{Physics of the
  Solid State} \textbf{2015}, \emph{57}, 700--708\relax
\mciteBstWouldAddEndPuncttrue
\mciteSetBstMidEndSepPunct{\mcitedefaultmidpunct}
{\mcitedefaultendpunct}{\mcitedefaultseppunct}\relax
\EndOfBibitem
\bibitem[Ishida \latin{et~al.}(1995)Ishida, Fujii, Kashiwagi, and
  Asano]{ishida1995}
Ishida,~S.; Fujii,~S.; Kashiwagi,~S.; Asano,~S. Search for half-metallic
  compounds in Co2MnZ (Z= IIIb, IVb, Vb element). \emph{Journal of the Physical
  Society of Japan} \textbf{1995}, \emph{64}, 2152--2157\relax
\mciteBstWouldAddEndPuncttrue
\mciteSetBstMidEndSepPunct{\mcitedefaultmidpunct}
{\mcitedefaultendpunct}{\mcitedefaultseppunct}\relax
\EndOfBibitem
\bibitem[Ze-Jin \latin{et~al.}(2017)Ze-Jin, Qing-He, Heng-Na, Ju-Xiang,
  Xian-Wei, and Zhi-Jun]{ze2017}
Ze-Jin,~Y.; Qing-He,~G.; Heng-Na,~X.; Ju-Xiang,~S.; Xian-Wei,~W.; Zhi-Jun,~X.
  Pressure-induced magnetic moment abnormal increase in Mn2FeAl and
  non-continuing decrease in Fe2MnAl via first principles. \emph{Scientific
  Reports} \textbf{2017}, \emph{7}, 16522\relax
\mciteBstWouldAddEndPuncttrue
\mciteSetBstMidEndSepPunct{\mcitedefaultmidpunct}
{\mcitedefaultendpunct}{\mcitedefaultseppunct}\relax
\EndOfBibitem
\bibitem[Deb and Sakurai(2000)Deb, and Sakurai]{deb2000}
Deb,~A.; Sakurai,~Y. Electronic structure of the Cu2MnAl Heusler alloy.
  \emph{Journal of Physics: Condensed Matter} \textbf{2000}, \emph{12},
  2997\relax
\mciteBstWouldAddEndPuncttrue
\mciteSetBstMidEndSepPunct{\mcitedefaultmidpunct}
{\mcitedefaultendpunct}{\mcitedefaultseppunct}\relax
\EndOfBibitem
\bibitem[Khvan \latin{et~al.}(2015)Khvan, Cheverikin, Dinsdale, Watson,
  Levchenko, and Zolotorevskiy]{khvan2015formation}
Khvan,~A.~V.; Cheverikin,~V.~V.; Dinsdale,~A.~T.; Watson,~A.; Levchenko,~V.~V.;
  Zolotorevskiy,~V.~S. Formation of metastable phases during solidification of
  Al--3.2 wt\% Mn. \emph{Journal of Alloys and Compounds} \textbf{2015},
  \emph{622}, 223--228\relax
\mciteBstWouldAddEndPuncttrue
\mciteSetBstMidEndSepPunct{\mcitedefaultmidpunct}
{\mcitedefaultendpunct}{\mcitedefaultseppunct}\relax
\EndOfBibitem
\bibitem[Xian \latin{et~al.}(2021)Xian, Peng, Zeng, Wang, and
  Gourlay]{xian2021al11mn4}
Xian,~J.; Peng,~L.; Zeng,~G.; Wang,~D.; Gourlay,~C. Al11Mn4 formation on Al8Mn5
  during the solidification and heat treatment of AZ-series magnesium alloys.
  \emph{Materialia} \textbf{2021}, \emph{19}, 101192\relax
\mciteBstWouldAddEndPuncttrue
\mciteSetBstMidEndSepPunct{\mcitedefaultmidpunct}
{\mcitedefaultendpunct}{\mcitedefaultseppunct}\relax
\EndOfBibitem
\bibitem[Reitz \latin{et~al.}(2024)Reitz, Hempelmann, Muller, Dronskowski, and
  Steinberg]{reitz2024bonding}
Reitz,~L.~S.; Hempelmann,~J.; Muller,~P.~C.; Dronskowski,~R.; Steinberg,~S.
  Bonding Analyses in the Broad Realm of Intermetallics: Understanding the Role
  of Chemical Bonding in the Design of Novel Materials. \emph{Chemistry of
  Materials} \textbf{2024}, \emph{36}, 6791--6804\relax
\mciteBstWouldAddEndPuncttrue
\mciteSetBstMidEndSepPunct{\mcitedefaultmidpunct}
{\mcitedefaultendpunct}{\mcitedefaultseppunct}\relax
\EndOfBibitem
\bibitem[Tolborg and Iversen(2021)Tolborg, and Iversen]{tolborg2021chemical}
Tolborg,~K.; Iversen,~B.~B. Chemical bonding origin of the thermoelectric power
  factor in half-Heusler semiconductors. \emph{Chem. Mater} \textbf{2021},
  \emph{33}, 5308--5316\relax
\mciteBstWouldAddEndPuncttrue
\mciteSetBstMidEndSepPunct{\mcitedefaultmidpunct}
{\mcitedefaultendpunct}{\mcitedefaultseppunct}\relax
\EndOfBibitem
\bibitem[Dronskowski(2004)]{dronskowski2004}
Dronskowski,~R. Itinerant ferromagnetism and antiferromagnetism from the
  perspective of chemical bonding. \emph{International journal of quantum
  chemistry} \textbf{2004}, \emph{96}, 89--94\relax
\mciteBstWouldAddEndPuncttrue
\mciteSetBstMidEndSepPunct{\mcitedefaultmidpunct}
{\mcitedefaultendpunct}{\mcitedefaultseppunct}\relax
\EndOfBibitem
\end{mcitethebibliography}

\end{document}